\documentclass[11pt,a4paper]{article}
\usepackage{jheppub}
\usepackage{epsfig}
\usepackage{latexsym}
\usepackage{eurosym}
\usepackage{setspace}
\usepackage{color}
\usepackage{amsmath}
\usepackage{amsfonts}
\usepackage{subcaption}
\usepackage{comment}
\usepackage{booktabs}
\usepackage{verbatim}
\usepackage{amssymb}
\usepackage{graphicx}

\providecommand{\norm}[1]{\lVert#1\rVert}
\providecommand{\U}[1]{\protect\rule{.1in}{.1in}}
\title{Analytic multi-Baryonic solutions in the $SU(N)$-Skyrme model at finite density}

\author[1,2]{Sergio L. Cacciatori}
\author[3]{Fabrizio Canfora}
\author[4]{Marcela Lagos}
\author[1,2]{Federica Muscolino}
\author[4]{Aldo Vera}

\affiliation[1]{Universit\`a  dell'Insubria, Dipartimento di Scienza ed Alta Tecnologia,\\
 via Valleggio 11, 22100, Como, Italy}
\affiliation[2]{INFN, via Celoria 16, 20133, Milano, Italy}
\affiliation[3]{Centro de Estudios Cient\'{\i}ficos (CECS), Casilla 1469, Valdivia, Chile}
\affiliation[4]{Instituto de Ciencias F\'isicas y Matem\'aticas, Universidad Austral de Chile,\\
Casilla 567, Valdivia, Chile}

\emailAdd{sergio.cacciatori@uninsubria.it}
\emailAdd{canfora@cecs.cl}
\emailAdd{marcela.lagos@uach.cl}
\emailAdd{federica.muscolino@uninsubria.it}
\emailAdd{aldo.vera@uach.cl}

\abstract{We construct explicit analytic solutions of the $SU(N)$-Skyrme model (for generic $N$) suitable to describe different phases of nuclear pasta at finite volume in $(3+1)$ dimensions. The first type are crystals of Baryonic
tubes (nuclear spaghetti) while the second type are smooth Baryonic layers (nuclear lasagna). Both, the ansatz for the spaghetti and the ansatz for the lasagna phases, reduce the complete set of Skyrme field equations to just one
integrable equation for the profile within sectors of arbitrary high topological charge. We compute explicitly the total energy of both configurations in terms of the flavor number, the density and the Baryonic
charge. Remarkably, our analytic results allow to compare explicitly the physical properties of nuclear spaghetti and lasagna phases. 
Our construction shows explicitly that, at lower densities, configurations
with $N=2$ light flavors are favored while, at higher densities,
configurations with $N=3$ are favored. Our construction also proves that in the high density regime (but
still well within the range of validity of the Skyrme model) the lasagna
configurations are favored while at low density the spaghetti configurations
are favored. Moreover, the integrability property
of the present configurations is not spoiled by the inclusion of the
subleading corrections to the Skyrme model arising in the 't Hooft expansion.
Finally, we briefly discuss the large $N$ limit of our configurations.}

\notoc

\begin{document}

\maketitle

\newpage
\tableofcontents

%%%%%%%%%%%%%%%%%%%%%%%%%%%%%%%%%%%%%%%%%%%%%%%%%%%%%%%%%%%%%%%%%%%%%

\section{Introduction}

%%%%%%%%%%%%%%%%%%%%%%%%%%%%%%%%%%%%%%%%%%%%%%%%%%%%%%%%%%%%%%%%%%%%%

One of the most fascinating phenomena appearing when a large amount of
Baryon charge is present within a finite volume (which has been confirmed
both phenomenologically and with numerical simulations) is the appearance of
ordered structures called \textit{nuclear pasta phase} (see \cite{pasta1}, 
\cite{pasta2}, \cite{pasta2a}, \cite{pasta2b}, \cite{pasta3}, \cite{pasta4}, 
\cite{pasta5}, \cite{pasta6}, \cite{pasta7}, \cite{pasta8}, \cite{pasta9}
and the nice up to date review \cite{pasta10}). Two of the most studied
shapes are \textit{nuclear spaghetti} (in which most of the Baryonic charge
lies within tube-shaped region) and \textit{nuclear lasagna} (in which most
of the Baryonic charge lies within layers of finite width). In the present
paper we will analyze in details these two types of nuclear pasta phase.

The properties of nuclear pasta attracted a lot of attention recently (see 
\cite{pasta4}, \cite{pasta5}, \cite{pasta6}, \cite{Schue}, \cite{Barros}, 
\cite{Acevedo} and references therein), but, due to both the very large
Baryon number typical of this phase as well as to the strong interactions
between them, only numerical results are available. Moreover, to obtain
meaningful numerical results is extremely challenging in these complex
systems and a very high computing power is required (as it can be deduced
from the references mentioned above). Needless to say, it is usually assumed
that any analytic approach is out of question. More in general, analytic
tools (different from perturbation theory) to analyze the phase diagram of
the low energy limit of Quantum Chromodynamics (QCD) at finite density and
low temperatures are extremely rare (especially due to the non-perturbative
nature of low energy QCD). This explains why the interesting but complex
phase diagram of QCD at finite density and low temperature is assumed to be
out of reach of analytic techniques (see \cite{newd3}, \cite{newd4}, \cite%
{newd5}, \cite{newd6} and references therein).

Nevertheless, the main aim of the present paper is \textit{to present an
analytic tool} suitable for the study of both, the nuclear spaghetti and
nuclear lasagna phases, within the Skyrme model \cite{skyrme}, which describes
the low energy limit of QCD at the leading order in the 't Hooft expansion 
\cite{Witten}, \cite{witten0}, \cite{bala0}, \cite{Bala1}, \cite{ANW} (see
also \cite{manton}, \cite{BaMa} and references therein). The Skyrme model is
a non-linear field theory for a scalar field $U$ taking values in the $SU(N)$
Lie group, being $N$ the flavor number. Despite the scalar nature of $U$,
the solitons of the theory are interpreted as Baryons.

Of course, one may ask: \textit{why should one insist so stubbornly in
finding analytic solutions if these equations can be solved numerically?} In
fact, the numerical tools in the references on nuclear pasta mentioned above
are powerful enough to shed new light on this phase. However, there are
indisputable reasons which compel us, whenever it is possible, to strive for
analytic solutions nevertheless.

\textit{First of all}, it could be enough to remind all the fundamental
ideas that the Schwarzschild and Kerr solutions in General Relativity and
the non-Abelian monopoles and instantons in Yang-Mills-Higgs theory
disclosed. Consequently, an analytic tool to study the nuclear pasta phase
can greatly enlarge our understanding of this complex phase. \textit{%
		Secondly and most importantly}, our analysis discloses relevant differences in
	the nuclear pasta phase arising from the competition between nuclear
	spaghetti and nuclear lasagna (which would have been hard to discover
	without the present analytic tools).

The methods introduced in \cite{56b}, \cite{56}, \cite{56b1}, \cite{56c}, 
\cite{crystal1}, \cite{crystal2}, \cite{crystal3}, \cite{crystal3.1}, \cite%
{crystal4}, \cite{firstube}, \cite{firsttube2} and \cite{gaugsksu(n)}
allowed the construction of several analytic and topologically non-trivial
solutions of the Skyrme model at finite Baryon density. As far as the
present paper is concerned, there are two relevant configurations analyzed
in those references. \textit{The first family} corresponds to ordered
Baryonic arrays in which (most of) the topological charge and total energy
are concentrated within tube-shaped regions\footnote{%
In \cite{Jackson} and \cite{nittain}, numerical string shaped solutions in
the Skyrme model with mass term have been constructed. However, those
configurations have a zero topological density (and they are expected to
decay into Pions). The configurations analyzed in the present paper are
topologically non-trivial and therefore can not decay into those of \cite%
{Jackson} and \cite{nittain}.}. \textit{The second family} corresponds to
configurations in which (most of) the topological charge and total energy
are concentrated within layers of finite width. Thus, the first family is
suitable to describe nuclear spaghetti while the second family to describe
nuclear lasagna. Indeed, on the nuclear spaghetti side, the similarity of
the contour plots in \cite{crystal1} with the spaghetti-like configurations
found (numerically) in the nuclear pasta phase (see the plots in \cite%
{pasta1}, \cite{pasta2}, \cite{pasta2a}, \cite{pasta2b}, \cite{pasta3} and 
\cite{pasta10}) is quite remarkable. On the nuclear lasagna side, the
contour plots of the energy density and Baryon density, which can be found
using the results in \cite{56c}, are very close to the numerical plots in 
\cite{pasta1}, \cite{pasta2}, \cite{pasta2a}, \cite{pasta2b}, \cite{pasta3},
for the lasagna case. Moreover, in \cite{gaugsksu(n)} the shear modulus of
lasagna configurations has been estimated analytically, the result being
close to recent numerical studies in \cite{pasta5} and \cite{pasta9}. These
results clearly show that the Skyrme model is suitable to describe the
nuclear pasta phase both qualitatively and quantitatively. With only two
exceptions in the lasagna case (namely, \cite{56b1} and\ \cite{gaugsksu(n)}%
), all the analytic results useful for the nuclear pasta phase (namely \cite%
{56}, \cite{56c}, \cite{crystal1}, \cite{crystal2}, \cite{crystal3.1}, \cite%
{crystal4}, \cite{firstube}, \cite{firsttube2}) have been obtained in the
case of the $SU(2)$-Skyrme model.

In the present manuscript we will extend the results of all the above
references to the case of the $SU(N)$-Skyrme model for generic $N$. This is
quite important from the viewpoint of the applications of the Skyrme model
since in many concrete situations (such as the nuclear pasta phase) it could
be relevant the inclusion of more flavors beyond Pions, Neutrons and
Protons: in particular, the most relevant cases (beyond $N=2$) are $N=3$ and 
$N=4$. Moreover, this generalization to generic $N$ also allows to use the
concept of non-embedded solutions (introduced in \cite{bala0} and \cite%
{Bala1}) which are solutions of the $SU(N)$-Skyrme model which cannot be
written as trivial embeddings of $SU(2)$ in $SU(N)$. Thus, combining the
strategy of \cite{crystal1}, \cite{crystal2}, \cite{crystal3}, \cite%
{crystal4}, \cite{firstube}, \cite{gaugsksu(n)}, with the generalization of
the Euler angles to $SU(N)$ of \cite{euler1}, \cite{euler2}, \cite{euler3},
we will construct non-embedded multi-Baryonic solutions of nuclear spaghetti
and nuclear lasagna.

The present analytic framework allows to write the explicit analytic
formulas for the energy density and the total energy of these configurations
for generic $N$, for large values of the Baryonic charge $B$ and for each
value of the size of the spatial volume within these configurations are
living. One can
compare the energy (seen as a function of the volume) of nuclear spaghetti
with the energy of nuclear lasagna \textit{for fixed values of the Baryonic
charge and fixed }$N$: the result is that in the high density regime (but
still well within the range of validity of the Skyrme model) the lasagna
configurations are favored while at low density the spaghetti configurations
are favored. In fact,
the comparison between the magnetic field decay of neutron stars and their
corresponding spin evolution obtained by numerical methods in references 
\cite{Nat1} and \cite{Nat2}, suggests that such structures exist.
In the light of the fact that lasagna and spaghetti phases are expected to have quite different
physical properties (see\ \cite{pasta10} and references therein).

\textit{A possible criticism} to the present analytic results is the
following: all these configurations have been constructed with a careful
choice of the ansatz for lasagna and spaghetti in the Skyrme case. In the 't
Hooft expansion, one should expect subleading corrections to the Skyrme
model (see, for instance, \cite{AdkinsNappi}, \cite{subleading1}, \cite%
{subleading2}, \cite{subleading3} and \cite{subleading4}) which may spoil
the present construction at subleading orders. However, the results in \cite%
{crystal3} strongly suggest that the same ansatz used in the present
manuscript \textit{allows to construct these configurations at any order in
the 't Hooft expansion analytically, no matter how many subleading terms are
included}.

The paper is organized as follows. In Section 2 we give a brief review of
the $SU(N)$-Skyrme model together with the general parameterization for the
fundamental fields. In Section 3 we construct analytical solutions
describing the nuclear spaghetti phase for generic values of $N$ and we
study its main features as the energy density distribution in terms of the
Baryonic charge, the flavor number and density. In Section 4 we first review
the nuclear lasagna phase and then we compare both configurations.
Also we briefly discuss the large $N$ limit of our configurations, the subleading corrections
and the inlcusion of a Isospin chemical potential. In the
final Section some conclusions and perspectives will be presented.

%%%%%%%%%%%%%%%%%%%%%%%%%%%%%%%%%%%%%%%%%%%%%%%%%%%%%%%%%%%%%%%%%%%%%

\section{The theory}

%%%%%%%%%%%%%%%%%%%%%%%%%%%%%%%%%%%%%%%%%%%%%%%%%%%%%%%%%%%%%%%%%%%%%

In this section we briefly review the $SU(N)$-Skyrme model that describes
the low energy limit of QCD at the leading order in the 't Hooft expansion
(see \cite{skyrme}, \cite{Witten}, \cite{witten0}, \cite{bala0}, \cite{Bala1}%
, \cite{ANW}, \cite{manton}, \cite{BaMa} and references therein) and we
present the general parameterization for the fields that will be used to
construct topological soliton solutions.

\subsection{The $SU(N)$-Skyrme model}

The action of the $SU(N)$-Skyrme model in $(3+1)$ dimensions is 
\begin{gather}  \label{I}
I= \int d^4x\sqrt{-g} \biggl[ \frac{K}{4}\text{Tr}\biggl( R_\mu R^\mu + 
\frac{\lambda}{8} F_{\mu\nu}F^{\mu\nu} \biggl) \biggl]\ , \\
R_\mu= U^{-1}\nabla_\mu U \ , \qquad F_{\mu\nu}=[R_\mu,R_\nu] \ , \qquad
U(x) \in SU(N) \ ,  \notag
\end{gather}
where $\nabla_\mu$ is the Levi-Civita covariant derivative, $K$ and $\lambda$
are positive coupling constants and $g$ is the metric determinant. In our
convention $c=\hbar=1$ and Greek indices $\{\mu, \nu, \rho, ...\}$ run over
the four dimensional space-time with mostly plus signature. Latin indices $%
\{i, j, k, ...\}$ are reserved for those of the internal space.

The Skyrme field $U$ is a map over the space-time taking values in the $%
SU(N) $ Lie group (being $N$ the flavor number), so that 
\begin{equation*}
R_\mu=R^i_\mu t_i \ ,
\end{equation*}
is in the $su(N)$ Lie algebra, where $t_i$ are the infinitesimal generators
of the $SU(N)$ group. We will see below that, for a given irreducible
representation of the group, it is possible to construct analytic Baryonic
configurations by deformations of embeddings of three dimensional Lie groups
into $SU(N)$.

The field equations of the model are obtained varying the action in Eq. %
\eqref{I} w.r.t. the $U$ field, 
\begin{equation}
\nabla^{\mu } \biggl( R_{\mu }+\frac{\lambda }{4}[R^{\nu},F_{\mu \nu }] %
\biggl) \ = \ 0\ ,  \label{Eqs}
\end{equation}
being these $(N^2-1)$ non-linear coupled second order differential equations.

The energy-momentum tensor, which is obtained using the standard formula 
\begin{equation*}
T_{\mu\nu} = -2 \frac{\partial \mathcal{L}}{\partial g^{\mu\nu}} +
g_{\mu\nu} \mathcal{L} \ ,
\end{equation*}
turns out to be 
\begin{equation}
T_{\mu \nu } \ = \ -\frac{K}{2}\text{Tr}\biggl(R_{\mu }R_{\nu }-\frac{1}{2}%
g_{\mu \nu }R_{\alpha }R^{\alpha }+\frac{\lambda }{4}(g^{\alpha \beta
}F_{\mu \alpha }F_{\nu \beta }-\frac{1}{4}g_{\mu \nu }F_{\alpha \beta
}F^{\alpha \beta })\biggl) \ .  \label{Tmunu}
\end{equation}
The topological charge is defined by 
\begin{gather}  \label{B}
B=\frac{1}{24\pi^{2}}\int_{\Sigma}\rho_{\text{B}}\ , \qquad \rho_{\text{B}%
}=\epsilon^{abc}\text{Tr}\biggl[\left( U^{-1}\partial_{a}U\right) \left(
U^{-1}\partial_{b}U\right) \left( U^{-1}\partial_{c}U\right) \biggl]\ ,
\end{gather}
where $\{a, b, c\}$ are spatial indices. When the topological charge density 
$\rho_{\text{B}}$ in Eq. \eqref{B} is integrated on a space-like surface, $B$
turns out to be the Baryonic number\footnote{%
We will see below that the Baryonic number for both nuclear pasta phases
depends explicitly on the flavor number $N$.}. Since we are interested in
describing states of Baryons, we need to impose that $\rho_{\text{B}}\neq 0$.

\subsection{General parameterization}

As one of our aims is the analysis of the finite density effects on the
multi-solitons, we need to put the system within a box of finite volume. The
simplest way to achieve this goal is to use the following flat metric 
\begin{equation}
ds^{2}=-dt^{2}+L_{r}^{2}dr^{2}+L_{\theta }^{2}d\theta ^{2}+L_{\phi
}^{2}d\phi ^{2}\ ,  \label{box}
\end{equation}%
where the adimensional spatial coordinates have the ranges 
\begin{equation}
0\leq r\leq 2\pi \ ,\quad 0\leq \theta \leq 2\pi \ ,\quad 0\leq \phi \leq
2\pi \ ,  \label{ranges}
\end{equation}%
so that the solitons are confined in a box of volume $V=(2\pi
)^{3}L_{r}L_{\theta }L_{\phi }$.
It is worth to emphasize here a relevant point: The ranges in Eq. \eqref{ranges} are enforced by the theory 
of Euler angles for $SU(N)$ \cite{euler1}, \cite{euler2}, \cite{euler3}. On the other hand, the constants $L_r$, $L_\theta$ and $L_\phi$
have dimensions of length. If one chooses physical unities such that $K=1$ and $\lambda=1$, then one would be measuring lengths in Fermi, $fm$.
The natural units of density is $\rho\sim 1/ (\text{volume})$. 

A remark is in order. While more or less everybody agrees on the value of the coupling constant $K$, the value of $\lambda$ is still under discussion. A typical
way to fix $\lambda$ is analyzing the properties of nucleons, as in \cite{ANW}. However, the size of the Skyrme coupling 
can change depending on whether one focuses on the properties of a single nucleon or of nuclear matter; see the discussion in \cite{BaMa}.

For the Skyrme field $U(x)\in SU(N)$ we use a parameterization in terms of
the generalized Euler angles \cite{euler1}, \cite{euler2}, \cite{euler3},
that is 
\begin{gather}
U=e^{\chi (x)\,(\vec{n}\cdot \vec{T})}\ ,  \label{U} \\
\vec{n}=(\sin {\Theta }\sin \Phi ,\sin \Theta \cos \Phi ,\cos \Theta )\ ,
\label{n}
\end{gather}%
where $\vec{T}=(T_{1},T_{2},T_{3})$ are three matrices of a given
representation of the Lie algebra $su(N)$, which will be chosen in order to
satisfy (see Appendix A for more details) the following relations 
\begin{equation*}
\lbrack T_{j},T_{k}]=\epsilon _{jkm}T_{m}\ ,\qquad \text{Tr}(T_{j}T_{k})=-%
\frac{N(N^{2}-1)}{12}\delta _{jk}\ .
\end{equation*}
In principle the functions $\chi $, $\Theta $, $\Phi $ that appear in the
ansatz in Eqs. \eqref{U} and \eqref{n} can depend on all the coordinates,
but (in the next section) we will choose these functions in such a way to
obtain analytical solutions with high topological charge. Then $\chi $ will
be identified as the soliton profile.

From Eqs. \eqref{B}, \eqref{U} and \eqref{n} it follows that the topological
charge density goes as 
\begin{equation}
\rho_{\text{B}} \ \sim \ (\sin^2(\frac{\chi}{2})\sin\Theta) \ d\chi \wedge d
\Theta \wedge d \Phi \ ,
\end{equation}
and therefore, as we want to consider only topologically non-trivial
configurations, we must demand that 
\begin{equation}  \label{notrho}
d\chi \wedge d \Theta \wedge d \Phi \ \neq \ 0 \ .
\end{equation}
It is important to note that Eq. \eqref{notrho} is a necessary but, in
general, not sufficient condition. In the next section we will show the
convenient form that the functions $\chi$, $\Theta$ and $\Phi$ should take
and the appropriate boundary conditions that lead to a non-vanishing
topological charge identified as the Baryonic number.

%%%%%%%%%%%%%%%%%%%%%%%%%%%%%%%%%%%%%%%%%%%%%%%%%%%%%%%%%%%%%%%%%%%%%

\section{Nuclear spaghetti phase}

%%%%%%%%%%%%%%%%%%%%%%%%%%%%%%%%%%%%%%%%%%%%%%%%%%%%%%%%%%%%%%%%%%%%%

In this section we will show that the $SU(N)$-Skyrme model admits analytical
solutions describing crystals of Baryonic tubes (nuclear spaghetti phase) at
finite volume.

\subsection{The ansatz} \label{1.1}

We need a good ansatz that respect the condition in Eq. \eqref{notrho} to
have a non-vanishing topological charge and also simplifies as much as
possible the field equations in Eq. \eqref{Eqs} in order to have analytic
solutions describing Baryonic states. According to Eq. \eqref{Eqs}, a good
set of conditions that allows to considerably simplify the field equations
are 
\begin{equation}
\nabla _{\mu }\Phi \nabla ^{\mu }\chi =\nabla _{\mu }\chi \nabla ^{\mu
}\Theta =\nabla _{\mu }\Phi \nabla ^{\mu }\Phi =\nabla _{\mu }\Theta \nabla
^{\mu }\Phi =\Box \Theta =\Box \Phi =0\ .  \label{conds}
\end{equation}%
Following the analysis in \cite{crystal1} and \cite{crystal2} one can see
that a suitable choice that satisfies both of the above criteria (specified
in Eqs. \eqref{notrho} and \eqref{conds}) is the following: 
\begin{gather}
\chi =\chi \left( r\right) \ ,\quad \Theta =q\theta \ ,\quad \Phi =p\left( 
\frac{t}{L_{\phi }}-\phi \right) \ ,  \label{ansatz} \\
q=\frac{1}{2}(2v+1)\ ,\quad v\in \mathbb{N}\ ,\quad p\neq 0\ .  \notag
\end{gather}

It is worth to note here that $p$ needs not to be an integer (as what is
important is that both $n$ and $np$ should be integer where $n$ is defined
in Eq. (\ref{Beq}) here below since $n$ is related to the number of
spaghetti in the box while $np$ is related to the Baryon charge). As the
analysis in the following sections will show, the energy-momentum tensor and
the Baryon density do not depend on the cooordinate $\phi $ while they do
depend on $r$ and $\theta $ (that's why these configurations describe
nuclear spaghetti). Thus, $p$ represents the Baryon number per unit of $%
L_{\phi }$. Hence, intuitively, values of $p$ less than 1 represents low
Baryon density per unit of length while values greater than 1 represents
high Baryon density per unit of length of spaghetti configurations. 

The above ansatz has allowed the construction of crystals of Baryonic tubes
as well as superconducting tubes in the $SU(2)$-Skyrme model \cite{crystal1}%
, \cite{crystal2}, \cite{crystal3}, \cite{crystal4}. Furthermore, the ansatz
in Eq. \eqref{ansatz} has a sort of universal character since it allows to
construct this kind of solutions also in the low energy limit of QCD \cite%
{crystal3}, making it clear that no matter how many subleading terms in the
't Hooft expansion are additionally considered in the Skyrme action, the
good properties of the above ansatz remain intact. That is the reason why we
will use the ansatz in Eqs. \eqref{box}, \eqref{U}, \eqref{n} and %
\eqref{ansatz} as a starting point for the construction of topological
solitons in the $SU(N)$ case.

On the other hand, the matrices $T_i$ define a three dimensional subalgebra
of $su(N)$ (see for example Appendix D of \cite{gaugsksu(n)}) and are given
explicitly as 
\begin{align}
T_1&=-\frac{i}{2}\sum_2 ^N \sqrt{(j-1)(N-j+1)}(E_{j-1,j}+E_{j,j-1}) \ ,
\label{T1} \\
T_2&=\frac{1}{2}\sum_2 ^N \sqrt{(j-1)(N-j+1)}(E_{j-1,j}+E_{j,j-1}) \ ,
\label{T2} \\
T_3&=i\sum_1 ^N (\frac{N+1}{2}-j)E_{j,j} \ ,  \label{T3}
\end{align}
with 
\begin{equation*}
(E_{i,j})_{mn}=\delta_{im}\delta_{jn}\ ,
\end{equation*}
being $\delta_{ij}$ the Kronecker delta (see Appendix A for more
mathematical details).

\subsection{Solving the system analytically}

It is a direct computation to verify that, according to the ansatz defined
in Eqs. \eqref{box}, \eqref{U}, \eqref{n} and \eqref{ansatz}, the components
of the tensor $R_{\mu }$ are 
\begin{align*}
R_{t}\ =& \ \frac{p}{L_{\phi }}(\sin {\chi }\tau_{3}+(1-\cos {\chi }%
)\tau_{2})\sin (q\theta )\ , \\
R_{r}\ =& \ \chi ^{\prime }\tau_{1}\ , \\
R_{\theta }\ =& \ q(\sin {\chi }\tau_{2}-(1-\cos {\chi })\tau_{3})\ , \\
R_{\phi }\ =& \ -L_{\phi }R_{t}\ ,
\end{align*}%
while the non-vanishing components of $F_{\mu \nu }$ turns out to be 
\begin{align*}
F_{tr}\ =& \ \frac{p}{L_{\phi }}(\sin {\chi }\tau_{2}-(1-\cos {\chi }%
)\tau_{3})\chi ^{\prime }\sin (q\theta )\ , \\
F_{t\theta }\ =& \ -\frac{2pq}{L_{\phi }}(1-\cos {\chi })\sin (q\theta
)\tau_{1}\ , \\
F_{r\theta }\ =& \ q(\sin {\chi }\tau_{3}+(1-\cos {\chi })\tau_{2})\chi
^{\prime }\ , \\
F_{r\phi }\ =& \ L_{\phi }F_{tr}\ , \\
F_{\theta \phi }\ =& \ L_{\phi }F_{t\phi }\ ,
\end{align*}%
where 
\begin{align*}
\tau_1=\vec{n}\cdot\vec{T}\ ,\qquad\tau_2=\partial_{\Theta}\tau_1\
,\qquad\tau_3=\frac{1}{\sin\Theta}\partial_{\Phi}\tau_1\ ,\qquad \left[%
\tau_i,\tau_j\right]={\varepsilon_{ij}}^k\tau_k\ .
\end{align*}
From the above, the $(N^2-1)$ coupled field equations of the Skyrme model in
Eq. \eqref{Eqs} with the ansatz defined in Eqs. \eqref{box}, \eqref{U} and %
\eqref{ansatz} are reduced to just one single ODE for the profile $\chi $,
namely 
\begin{equation}
\biggl(1+\frac{\lambda q^{2}}{L_{\theta }^{2}}\sin ^{2}(\frac{\chi }{2})%
\biggl)\chi ^{\prime \prime }-\frac{q^{2}L_{r}^{2}}{L_{\theta }^{2}}\biggl(1-%
\frac{\lambda }{4L_{r}^{2}}\chi ^{\prime 2}\biggl)\sin (\chi )\ =\ 0\ .
\label{Eqchi}
\end{equation}
Note that Eq. \eqref{Eqchi} does not depend on $N$, so that the Skyrme
equation is independent of the Lie group under consideration\footnote{%
We will see below that, although the profile does not depend on $N$, both
the energy and the Baryonic charge are functions of $N$, as expected.}.
Futhermore, the above equation can be reduced to the following first order
ODE 
\begin{equation}
\biggl(1+\frac{q^{2}\lambda }{L_{\theta }^{2}}\sin ^{2}(\frac{\chi }{2})%
\biggl)\chi ^{\prime 2}+\frac{2L_{r}^{2}q^{2}}{L_{\theta }^{2}}\cos (\chi
)=E_{0}\ ,  \label{dchi}
\end{equation}%
where $E_{0}$ is an integration constant. Eq. \eqref{dchi} is explicitly
solvable in terms of generalized Elliptic Integrals \cite{Elliptics} and it
is reducible to the following quadrature%
\begin{gather}  \label{qua}
\frac{d\chi }{\eta (\chi ,E_{0})}=\pm dr\ , \qquad \eta (\chi ,E_{0})=\pm %
\biggl[\frac{E_{0}L_{\theta }^{2}-2L_{r}^{2}q^{2}\cos (\chi )}{L_{\theta
}^{2}+q^{2}\lambda \sin ^{2}(\frac{\chi }{2})}\biggl]^{\frac{1}{2}}\ .
\end{gather}%
The integration constant $E_{0}$ plays a fundamental role in determining the
Baryonic charge, as we will see here below.

\subsection{A constraint from stability}

When the field equations reduce to a single equation for the profile in a
topologically non-trivial sector (as in the present case) one says that ``%
\textit{the hedgehog property holds}". Often (although not always, see \cite%
{shifman1}, \cite{shifman2} and references therein) the most dangerous
perturbations\footnote{%
Namely, a perturbation which could lead to a decrease in the energy of the
system.} are those perturbations of the profile which keep the hedgehog
property.

In the present case, these dangerous perturbations are of the following form:%
\begin{equation}
\chi \rightarrow \chi +\varepsilon \xi \left( r\right) \ ,\ \ \ \left\vert
\varepsilon \right\vert \ll 1\ ,  \label{pert}
\end{equation}%
which do not change the $SU(N)$ Isospin degrees of freedom but only the
profile. It is a direct computation to show that the linearized field
equations under the perturbation in Eq. (\ref{pert}) always has the
following zero-mode: $\xi \left( r\right) =\partial _{r}\chi \left( r\right) 
$, where $\chi $ is the solution of the field equations (satisfying the
boundary conditions defined here below). From this we can deduce a
constraint (which is a necessary condition for stability) on the integration
constant $E_{0}$ in Eqs. (\ref{dchi}) and \eqref{qua}, namely 
\begin{equation*}
E_{0}>\frac{2L_{r}^{2}q^{2}}{L_{\theta }^{2}}.
\end{equation*}%
If the above condition is satisfied the zero mode $\xi \left( r\right)
=\partial _{r}\chi \left( r\right) $ has no node (since $\partial _{r}\chi
\left( r\right) $ does not vanish in this case) and so the system is stable
under these perturbations.

\subsection{Boundary conditions and Baryonic charge} %\textcolor{red}{(aggiungere commento di Sergio)}}

From Eq. \eqref{B} and using Eqs. \eqref{U}, \eqref{n} and \eqref{ansatz} we
can compute the topological charge density of the configurations presented
above, which turns out to be 
\begin{gather}
\rho_{\text{B}} \ = \ N(N^2-1) \ p q \sin(q\theta)\sin^2\biggl(\frac{\chi}{2}%
\biggl)\chi^{\prime} \ .  \label{measure}
\end{gather}
We see that, in fact, the topological charge density depends on the Lie
group through the factor $N(N^2-1)$. Integrating the above over a space-like
hypersurface in the ranges defined in Eq. \eqref{ranges}, we arrive to the
following expression for the topological charge 
\begin{align}  \label{Beq}
B \ = & \ 2 n p \frac{N(N^2-1)}{12} \ ,
\end{align}
where we have used the following boundary conditions 
\begin{align}  \label{BC}
\chi(0)=0 \ , \quad \chi(2\pi) =& 2n\pi \ ,
\end{align}
with $n$ an integer and $q$ specified in Eq. \eqref{ansatz}. These
conditions arises if we require the $U$ field to cover an entire cycle in
the range of the coordinates in Eq. \eqref{ranges}. As shown in \cite{euler2}%
, this is accomplished imposing that the variables explicitly appearing in
the ``measure'' in Eq. (\ref{measure}) must run in a range where the measure
is non-vanishing. This fact immediately implies that $\theta \in [0,\pi/q]$
(for a fundamental solution) and $\chi\in[0,2\pi]$. Therefore the
topological charge for the spaghetti phase depends on $N $ and it is labeled
by the integer $n$ that appears in the boundary conditions in Eq. \eqref{BC}
and the value of $p$ in the ansatz in Eq. \eqref{ansatz}. Note that the
integration constant $E_{0}$ in Eq. \eqref{qua} is fixed in terms of $n$
through the equation 
\begin{equation}
n \int _0^{2\pi} \frac{1}{\eta(\chi,E_0)}d\chi = 2\pi \ ,  \label{E0n}
\end{equation}%
that will always have a real solution.\\
We want to remark here that, despite our choice of the ranges may look to define periodic boundary conditions, it is not the case. The boundary condition are chosen so that the map embedding the spatial rectangle into the $SU(N)$ exactly wraps a cycle in 
$H_3(SU(N),\mathbb Z)$. This is a topological condition necessary to have a non vanishing Baryon number, while it is easy to see that periodic boundary conditions in all variables would lead to a vanishing Baryon number (see \cite{euler1,euler2}).

\subsection{Characterizing the $SU(N)$ nuclear spaghetti phase}

At this point it is important to emphasize that using the ansatz introduced
in Section \ref{1.1} we have reduced the complete set of Skyrme equations to
just one equation in Eq. \eqref{Eqchi} for the profile $\chi$. Even more,
this equation can be solved analytically and does not depends on $N$.

In what follows we will set the values of the coupling constants as $K=2$
and $\lambda=1$ for the numerical computations. We will also define the
density in the case $L_r=L_\theta=L_\phi \equiv L$, as $\rho=1/(2 \pi L)^3$.

\begin{figure}[h!]
\centering
\includegraphics[scale=.8]{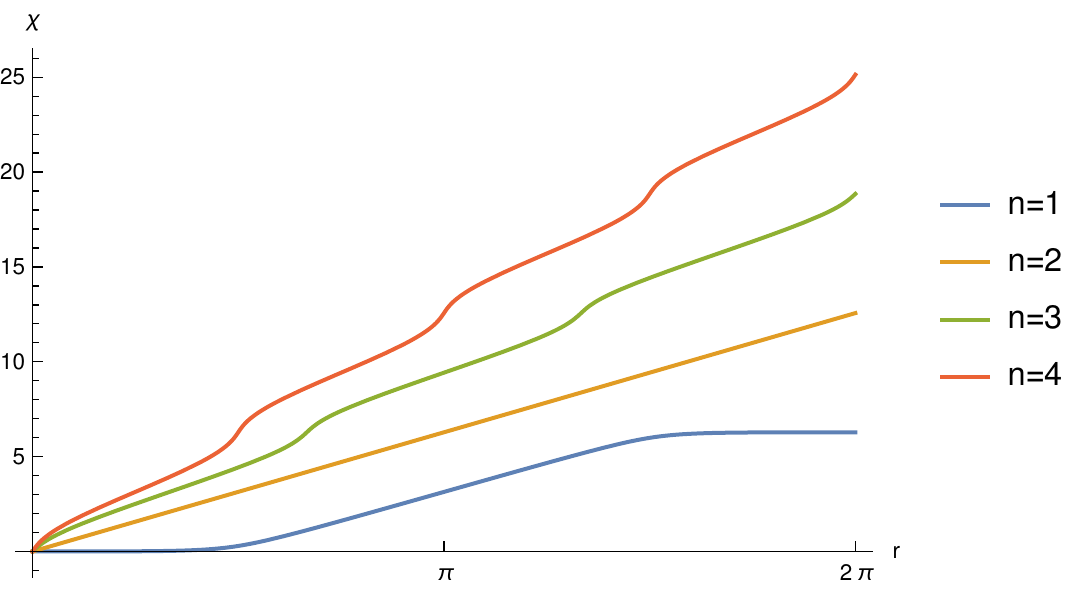}
\caption{Soliton profile $\protect\chi=\protect\chi(r)$ for different values
of $n$, with $q=\frac{9}{2}$ and $L=1$.}
\label{Fig1}
\end{figure}

\begin{figure}[h!]
\centering
\includegraphics[scale=.35]{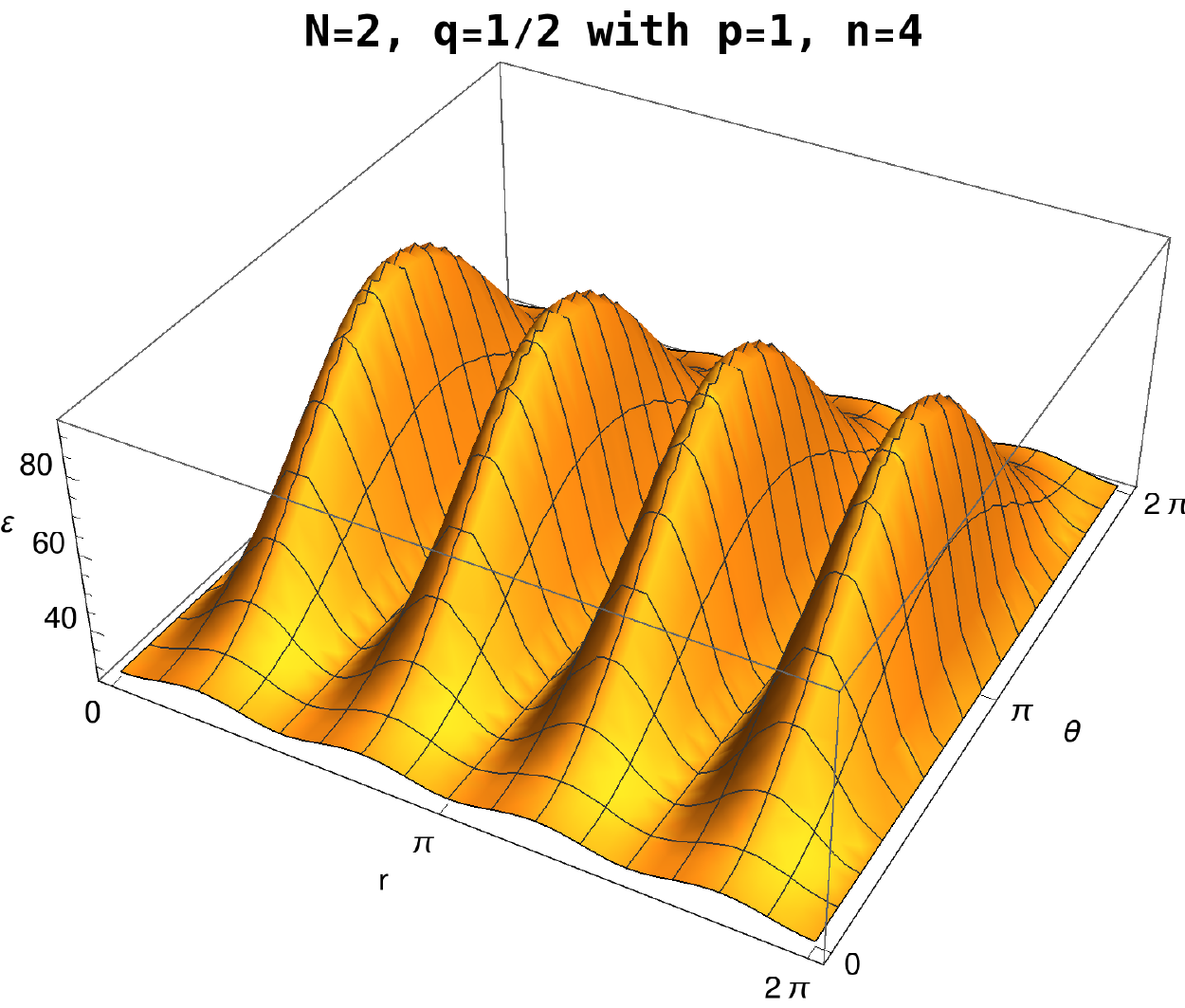} \qquad %
\includegraphics[scale=.35]{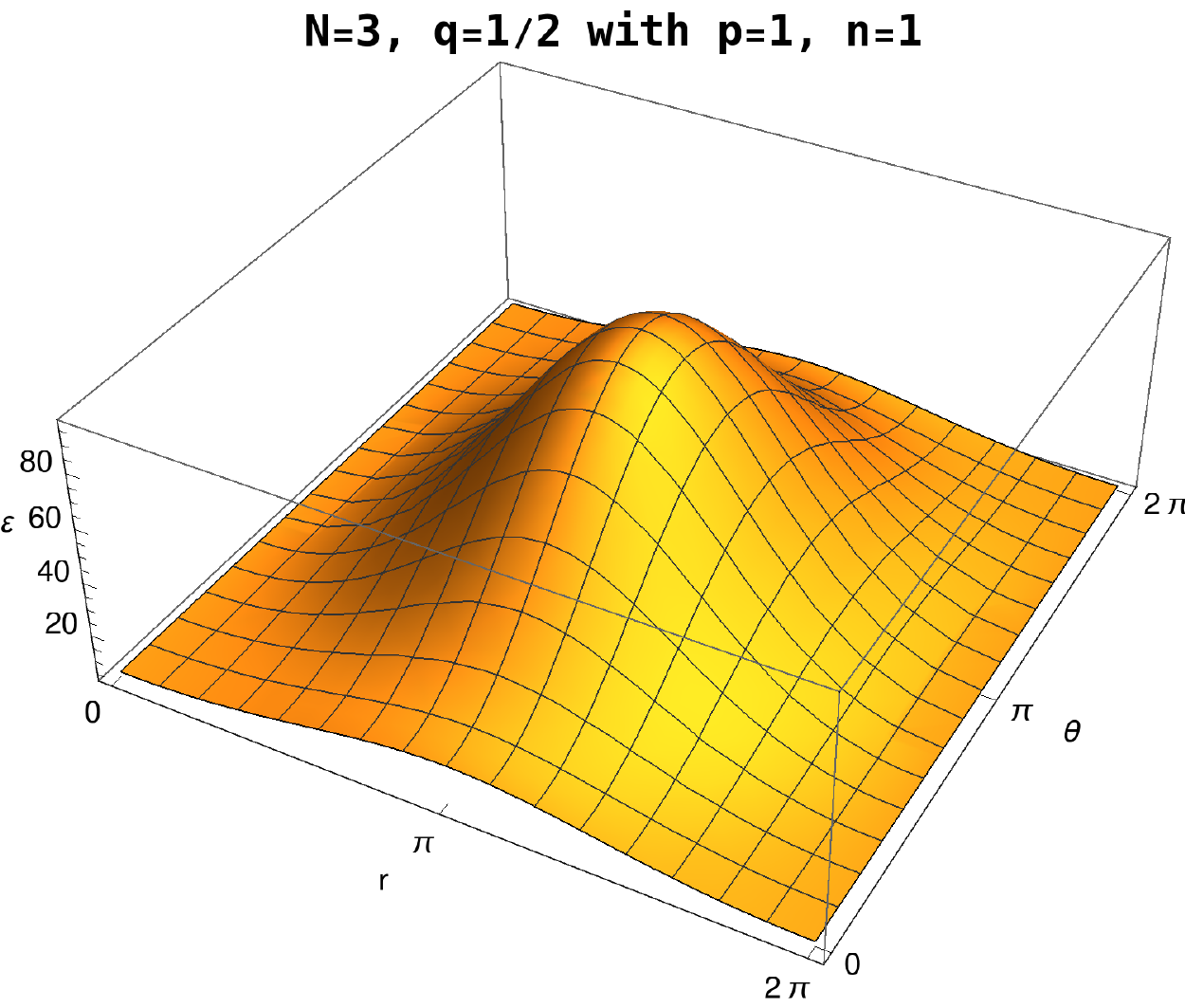} \qquad \qquad %
\includegraphics[scale=.35]{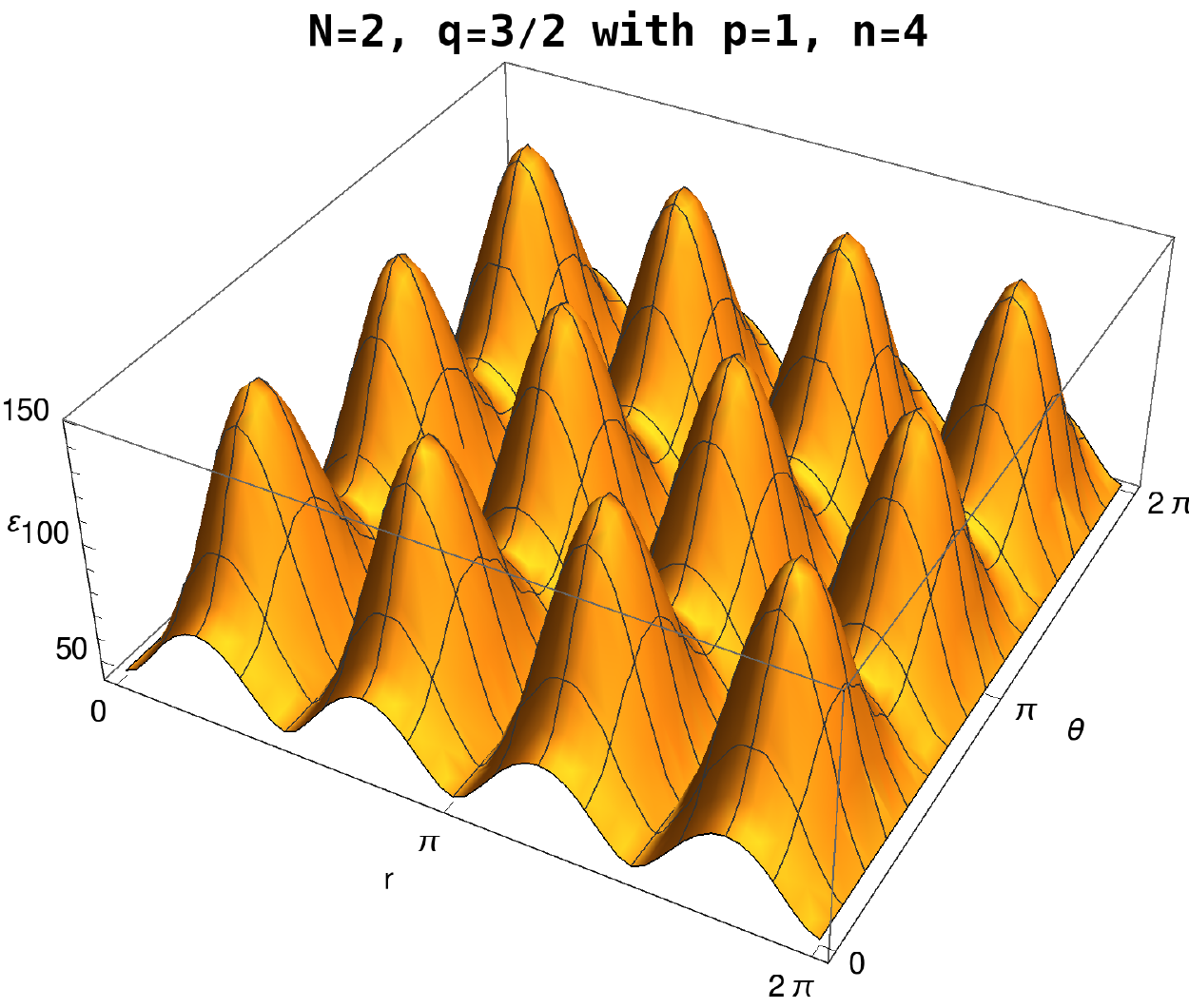} \qquad %
\includegraphics[scale=.35]{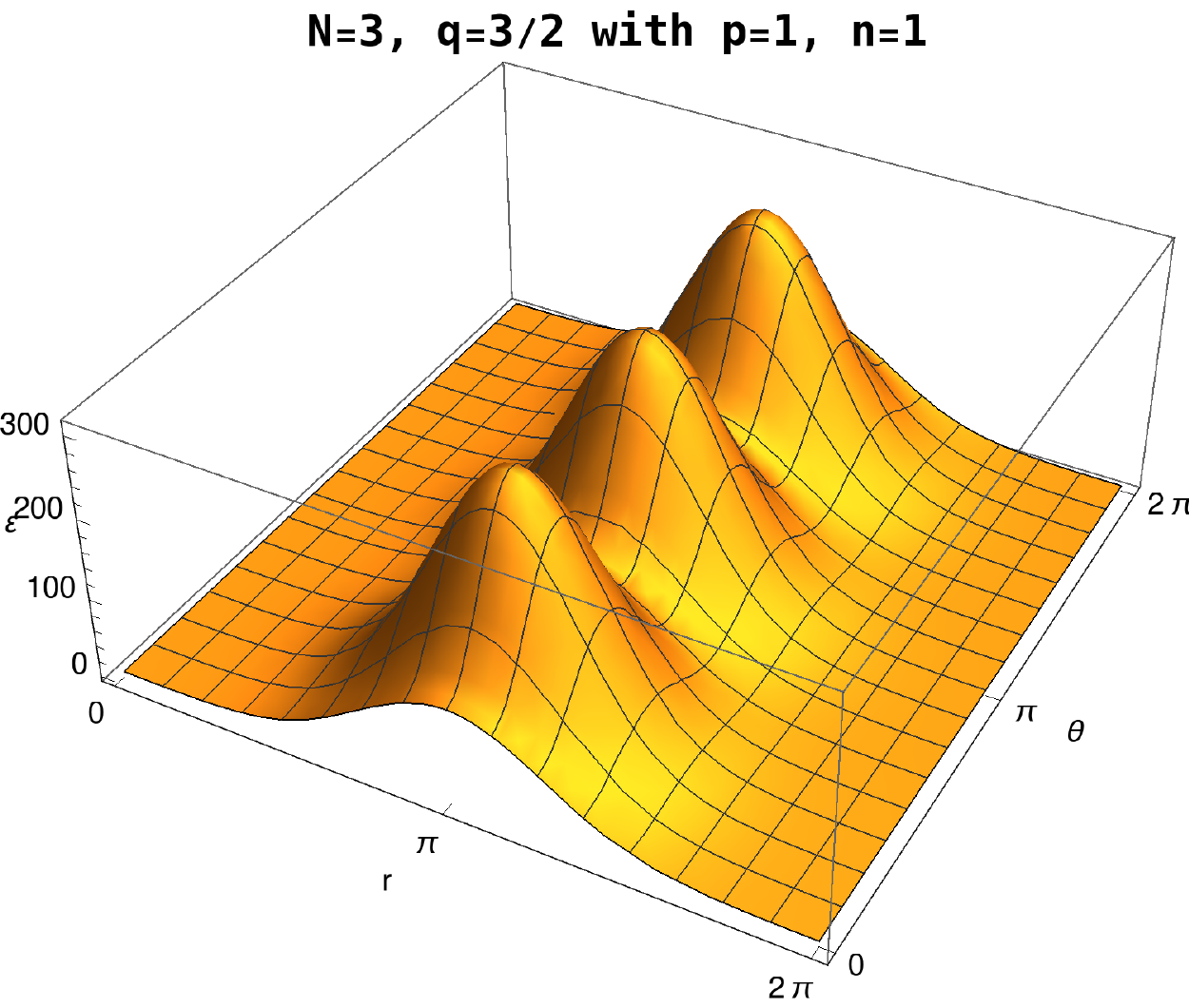}
\caption{Energy density for some of the allowed configurations with $B=4$
for different values of $q$ and with $p=1$ and $L=1$.}
\label{Fig2}
\end{figure}

Fig. \ref{Fig1} shows the behavior of the profile of the soliton
configurations as a function of the parameter $n$, which determines
different values of the Baryonic charge according to Eq. \eqref{Beq}. 
Now, even though the Skyme equation does not depend on $N$ explicitly (see
Eq. \eqref{Eqchi}), the energy density does. In fact, according to Eq. %
\eqref{Tmunu} and using Eqs. \eqref{box}, \eqref{U} and \eqref{ansatz}, the
energy density of these configurations is 
\begin{equation}
\mathcal{E} \ = \ \frac{K}{4}\frac{N(N^2-1)p}{12 L_r L_\theta L_\phi^2 }
\left( \rho_0 + 2\sin^2(q\theta) \rho_1 \right) \ ,  \label{3.14}
\end{equation}
where the functions $\rho_0$ and $\rho_1$ are given respectively by 
\begin{align*}
\rho_0 \ = \ & \frac{L_\phi^2}{p}\biggl[ 4L_r^2q^2 \sin^2(\frac{\chi}{2}) +%
\biggl(L_\theta^2+q^2 \lambda \sin^2(\frac{\chi}{2})\biggl)\chi^{\prime 2 }%
\biggl] \ , \\
\rho_1 \ = \ & p\sin^2(\frac{\chi}{2}) \biggl[ 4L_r^2\biggl( %
L_\theta^2+q^2\lambda \sin^2(\frac{\chi}{2}) \biggl) +L_\theta^2 \lambda
\chi^{\prime 2 }\biggl] \ .
\end{align*}
In Fig. \ref{Fig2} we show plots of the energy density for some of the
allowed spaghetti configurations with $B=4$, which is the lowest value of
the topological charge in both $N=2$ and $N=3$ cases according to Eq. %
\eqref{Beq} (see Table \ref{TableB} in Appendix B for the explicit values of 
$B=B(N)$) when both, the value of the parameters $q$ and $n$ increase. We
see that the number of peaks in the $r$ direction increases as $n$
increases, while the value of the parameter $q$ repeats the pattern in the $%
\theta$ direction of the lattice in which the solitons are confined. 

\begin{figure}[h!]
\centering
\includegraphics[scale=.65]{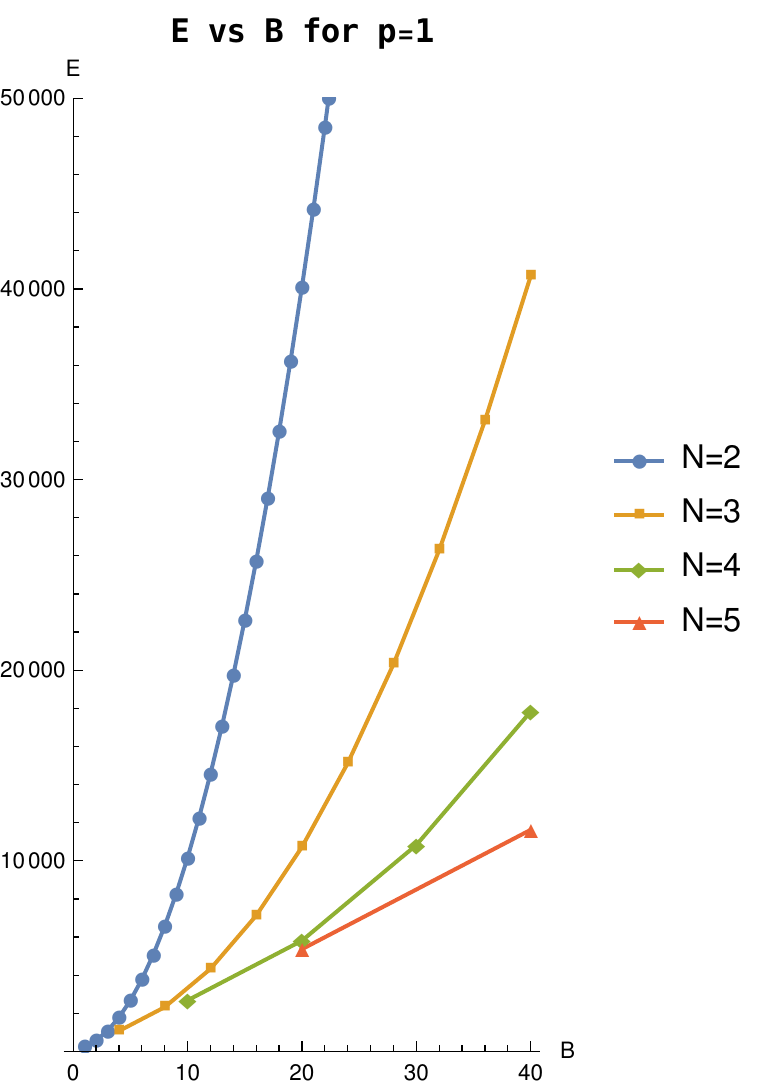} \qquad %
\includegraphics[scale=.65]{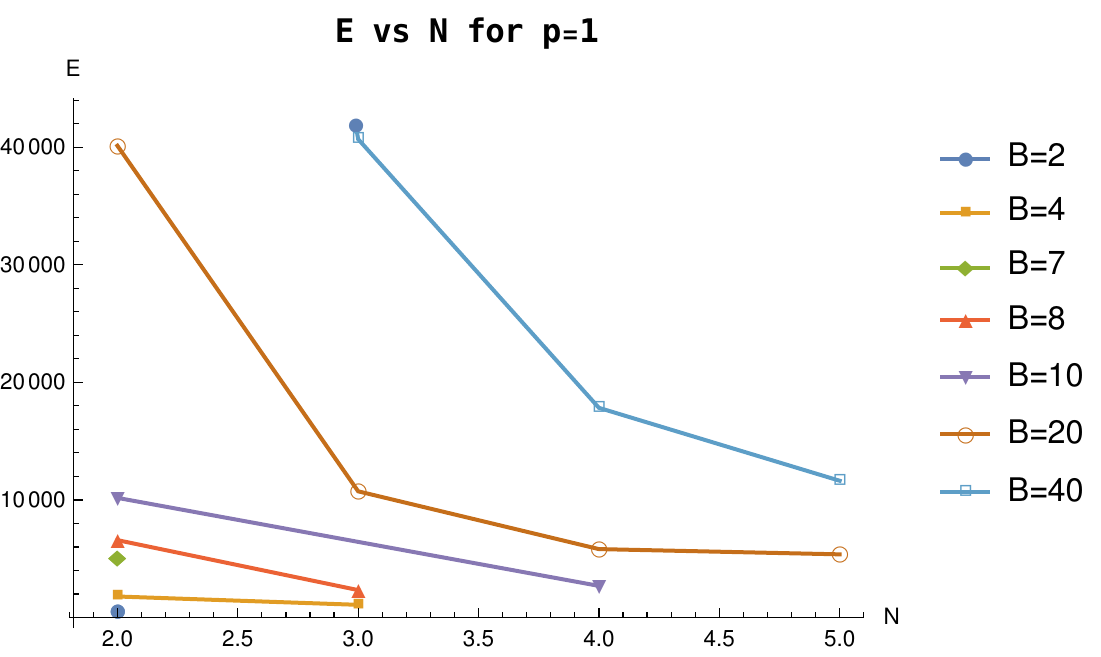}
\caption{Left: Energy versus Baryonic charge for different values of $N$.
Right: Energy versus $N$ for different values of $B$. Here $p=1$, $q=\frac{1%
}{2}$ and $L=1$.}
\label{Fig3}
\end{figure}

In Fig. \ref{Fig3}, in the left side, we have plotted the total energy as a
function of the Baryonic charge for different fixed values of the flavor
number. In the right side we show the behavior of the total energy in terms
of the flavor number once fixed the topological charge. From here we can see
two interesting facts: when the value of $N$ is fixed the energy is an
increasing function of $B$. In the opposite way, when a particular value of $%
B$ is chosen, the energy decreases with $N$. This means that if we consider
a fixed volume box, as we add more Baryons to the box the energy of the
system increases, which is the expected result due to the repulsion energy
between Baryons. Now, if we compare the same fixed volume box containing the
same Baryonic number but for different values of the group dimension, as we
increase $N$ the energy of the configuration will be lower. According to the
above, for example, the $4$-Baryon state for the $SU(3)$ group in Fig. \ref%
{Fig2} (up-right) is less energetic than the four independent Baryons state
for the $SU(2)$ group in Fig. \ref{Fig2} (up-left). The amount of energy per
Baryon has the same behavior. 

\begin{figure}[th]
\centering
\includegraphics[scale=.7]{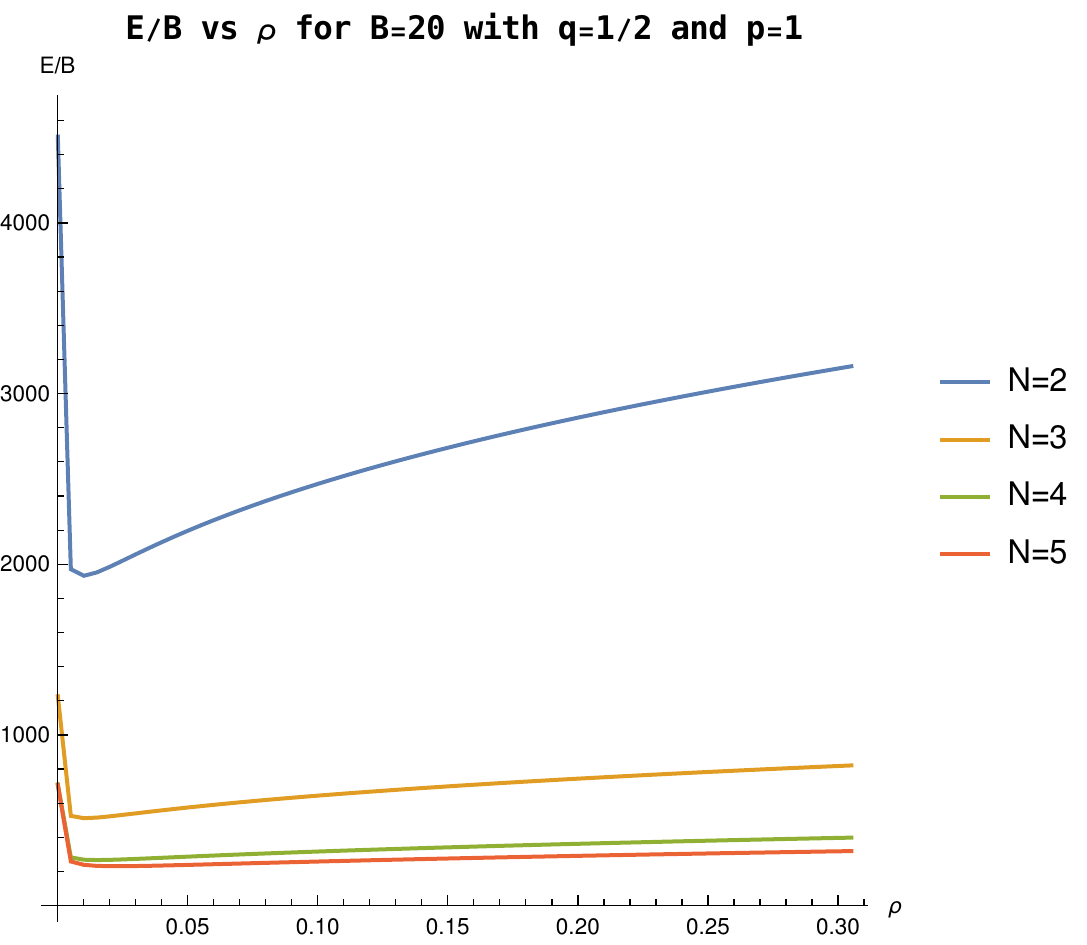}
\caption{Energy per Baryonic charge as a function of the density with $q=%
\frac{1}{2}$. We show all the allowed configurations with $B=20$ and $p=1$.
One can see that the behavior of the curves has the characteristic
``u-shape'' of nuclear pasta shown in \protect\cite{pasta10}.}
\label{Fig4}
\end{figure}

In Fig. \ref{Fig4} we show the behavior of the energy as the density
changes; this for fixed values of $B$ and $N$ and considering a cubic box in
which the soltions are confined. We can see that in the high density sector
the energy of the system increases, being the divergence at the end expected
since at low scales the Skyrme model (which is an effective model of Baryons
and Pions) should be replaced by QCD. However, there is a critical point
from which the behavior reverses in such a way that at low density the
energy of the system becomes a decreasing function of the density. This
``u-shaped" behavior of the energy versus density that we can see from Fig. %
\ref{Fig4} is in accordance with what has been obtained in numerical
simulations of nuclear pasta (see \cite{pasta10}). In fact, the above is one
of the main goals of the present work: relevant features of the nuclear
pasta state, that until now has only been possible to study numerically, can
be characterized from analytical solutions of the $SU(N)$-Skyrme model.

\begin{figure}[th]
\centering
\includegraphics[scale=.495]{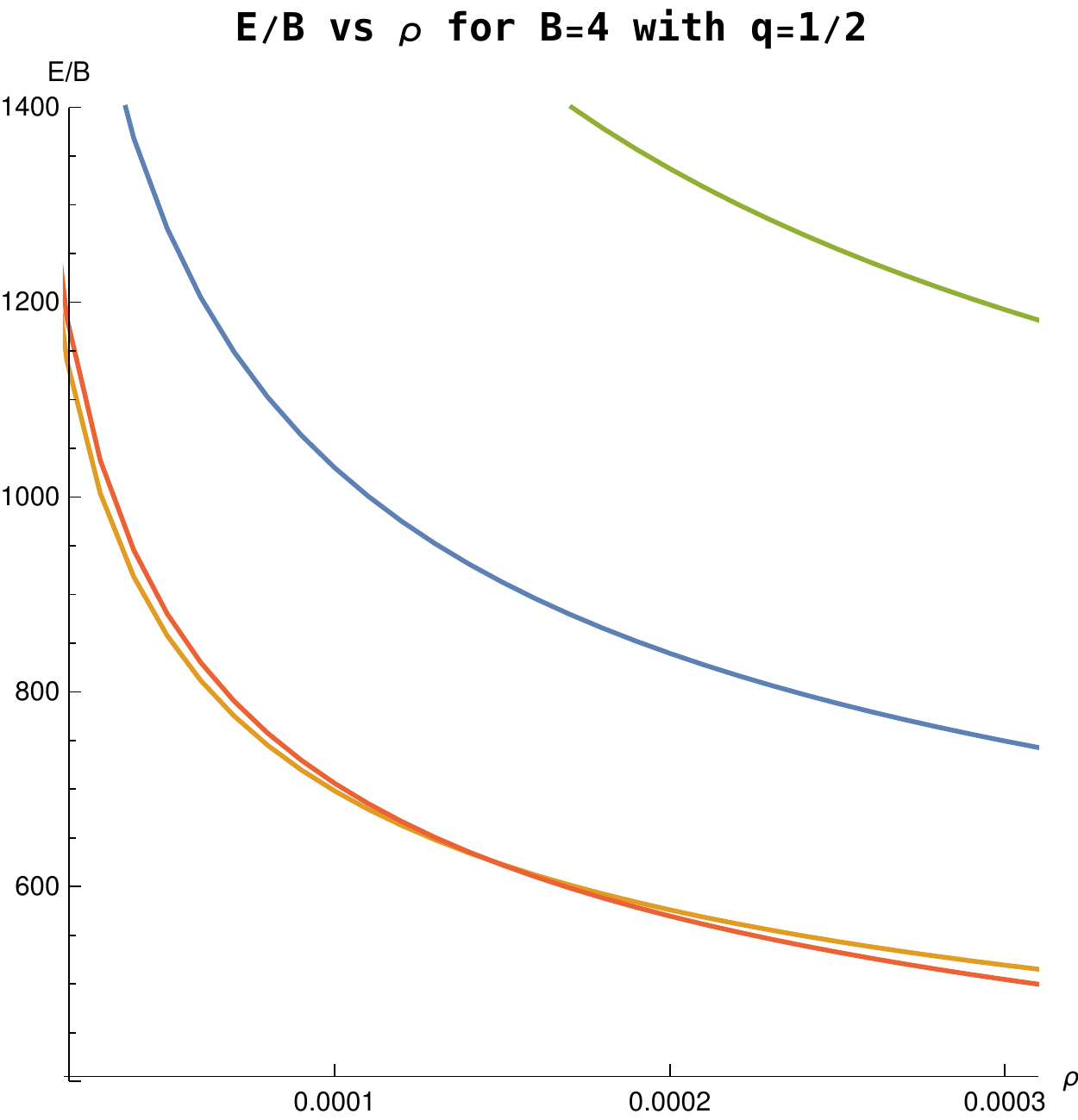}\quad %
\includegraphics[scale=.68]{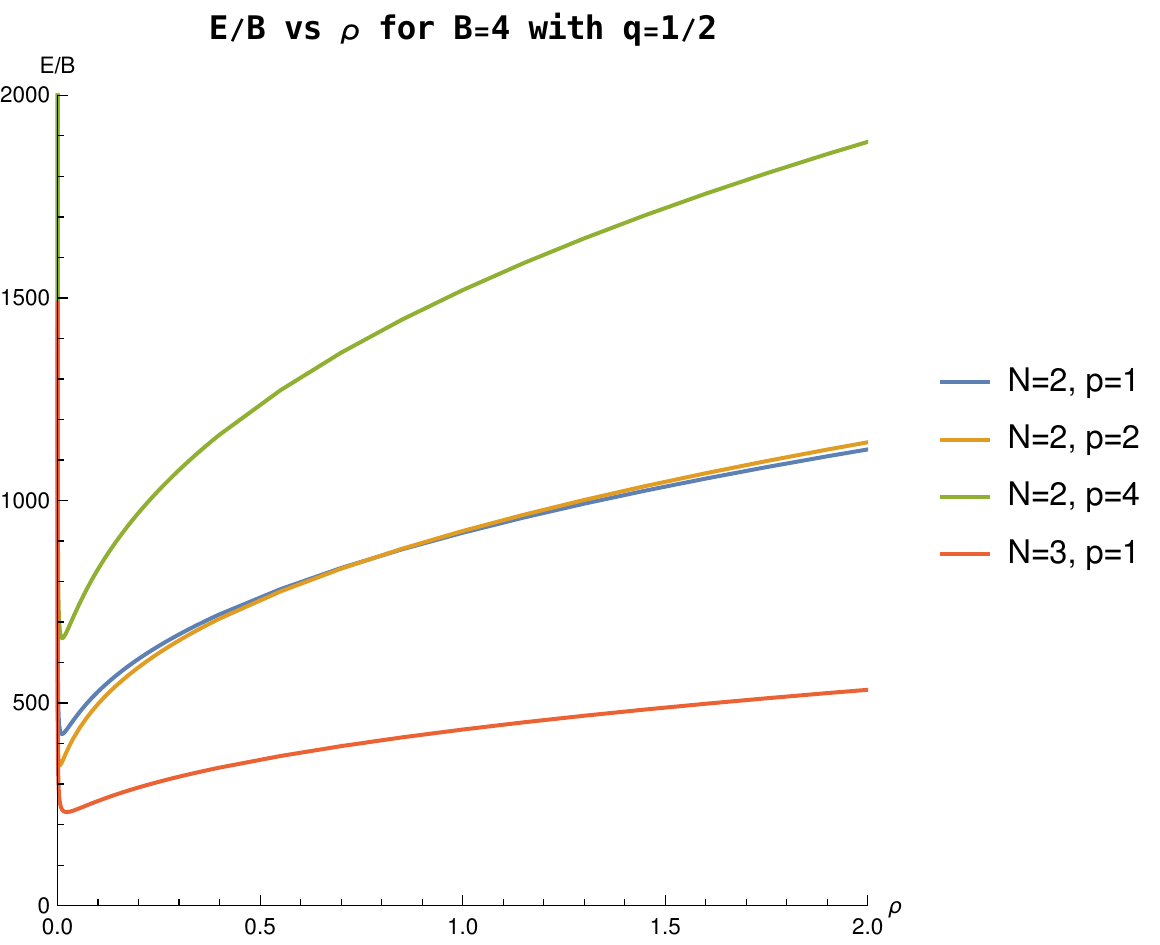}
\caption{Energy per Baryonic charge as a function of the density with $q=%
\frac{1}{2}$. For $B=4$ we see (when $N$, $p$ and $n$
varies) left: At low densities the energetically preferable spaghetti
configuration in $N=2$, whereas at larger densities the configuration with $%
N=3$ has the lowest energy. Right: We see a transition that occurs between
configurations with $N=2$, for different values of $n$ and $p$.}
\label{Fig4.1}
\end{figure}

Fig. \ref{Fig4.1} shows that, for different configurations of nuclear
spaghetti with the same Baryonic charge, there are transitions that depend
on the values of the constants $p$ and $n$ that define the topological
charge according to Eq. \eqref{Beq}. In particular,
from the left plot we see that at low densities the energetically preferable
spaghetti configuration is the one with two flavors, whereas at larger
densities the configuration with three flavors has the lowest energy. Also,
from the right plot and for larger densities we can see a transition between
spaghetti configurations within the same internal group ($N=2$), for
different values of $p$ and $n$. 

The following comment is in order. Traditionally (see, for instance, \cite{ANW}) the Skyrme coupling constant is fixed by requiring the best possible agreement with the static properties of the Neutron and $\Delta_{++}$. However (while everybody agrees 
on the Pions coupling constant K) there is no common agreement yet on the value of the Skyrme coupling constant $\lambda$. In the plot in Fig. \ref{Fig4.1} we have used the ``traditional value'' for the Skyrme coupling constant $\lambda$. On the other hand, 
it could be convenient to fix 
$\lambda$ in order to get an excellent description of nuclear pasta. We think that the present results strongly support this point of view.

\begin{figure}[ht]
\centering
\includegraphics[scale=.6]{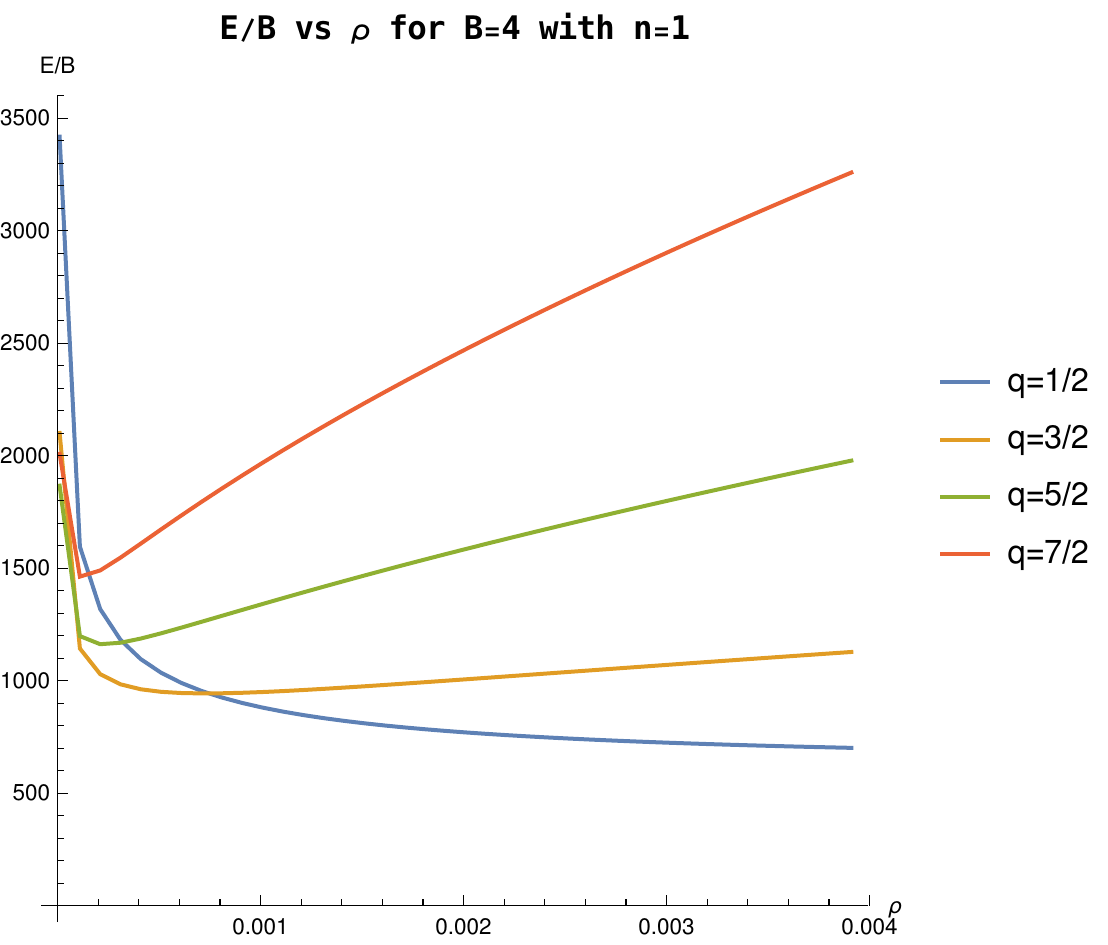}
\caption{Energy per Baryonic charge as function of the density with $B=4$ in 
$SU(2)$ and for different values of $q$. We see that, at high density,
configurations with lower values of $q$ are energetically favoured while, in
the low density sector, the configurations with higher values of $q$ are the
favoured ones. }
\label{FigqS}
\end{figure}

Finally, there is also a transition that appears
when the value of the $q$ parameter varies. In fact, from Fig. \ref{FigqS} we can see
that as the density decreases the configurations with higher values of $q$
becomes the energetically favored ones.

%%%%%%%%%%%%%%%%%%%%%%%%%%%%%%%%%%%%%%%%%%%%%%%%%%%%%%%%%%%%%%%%%%%%%

\section{Nuclear spaghetti versus nuclear lasagna}

%%%%%%%%%%%%%%%%%%%%%%%%%%%%%%%%%%%%%%%%%%%%%%%%%%%%%%%%%%%%%%%%%%%%%

In this section we will compare the nuclear
spaghetti and nuclear lasagna phases.

\subsection{Review of the $SU(N)$ nuclear lasagna phase}

Here we will summarize the most important features of the nuclear lasagna
solutions that were previously constructed in \cite{gaugsksu(n)}, in order
to compare these solutions with the new nuclear spaghetti phase presented in
the previous section of this manuscript.

The ansatz for the Skyrme field that allows to construct the analytic
nuclear lasagna phase, as a solution of the $SU(N)$-Skyrme model in Eqs. %
\eqref{I} and \eqref{Eqs}, is given by 
\begin{equation}
U_{L}=e^{\sigma \Phi k}e^{h(r)}e^{m\theta k}\ ,  \label{U2}
\end{equation}%
where 
\begin{equation}
\sigma =2^{-\frac{(-1)^{N}+1}{2}}\ ,\qquad \Phi =\frac{t}{L_{\phi }}-\phi \
,\qquad k=\sum_{j=1}^{N-1}\biggl(c_{j}E_{j,j+1}-c_{j}^{\ast }E_{j+1,j}\biggl)%
\ .  \label{U2b}
\end{equation}%
Here $m$ is a non-vanishing integer number and $c_{j}$ are arbitrary complex
numbers with the constraint that $e^{\theta k}$ must be periodic with period 
$2\pi $. The last is indeed a highly nontrivial constraint, which has been
solved in \cite{gaugsksu(n)}. We will consider again the metric of a box in
Eq. \eqref{box} with the same ranges in the coordinates.

It is worth to remark that in \cite{gaugsksu(n)} it has been shown that the solutions of such constraints define a moduli space (the set of allowed $c_j$) which becomes larger and larger with $N$. For a generic choice of $c_j$, and thus of $k$, in such moduli
space, the linear space generated by the matrices $h(r), k$ and $[h(r),k]$ is not a tridimensional subalgebra of $su(N)$. This happens only by choosing the parameters $c_j$ in a subset of vanishing measure of the moduli space. This means that for a 
generic choice of the parameters, despite expression (\ref{U2}) has the form of an Euler parametrization of $SU(2)$, it does not define a subgroup of $SU(N)$, but just a submanifold. In this sense the Euler construction is not an embedding of $SU(2)$ into 
$SU(N)$ as Lie groups, but only as manifolds. Remarkably, the very specific choices that give rise to embeddings of $SU(2)$ in $SU(N)$ (and other simple Lie groups) have been determined by E. Dynkin in \cite{Dy-57}. In the very particular cases 
(which are instead the general case for the spagetti construction) when 
the choice of the parameters defines a subgroup, then we have an embedding which is called non-trivial if the image contains sub representations of spin different from $0$ and $1/2$, and trivial otherwise. In the non trivial case, one usually says that the
corresponding solutions of the Skyrme equations are of true $SU(N)$ type (see \cite{gaugsksu(n)} and references therein).\newline
It can be directly checked that the configurations described by the ansatz
in Eqs. \eqref{box} and \eqref{U2} reduce the complete set of Skyrme field
equation (see \cite{gaugsksu(n)} for details) to the following ODE: 
\begin{equation}
h^{\prime \prime }=\frac{\lambda q^{2}}{4L_{\gamma }^{2}}\left(
[k,[k,h^{\prime \prime }]]-[k,[h^{\prime },[h^{\prime },k]]]\right) \ .
\label{Eqh}
\end{equation}%
Eq. \eqref{Eqh} can be directly solved, and its solution is given by 
\begin{equation}
h(r)=\frac{1}{2}rv_{\varepsilon }\ ,\qquad v_{\varepsilon }=\sum_{j,k}{%
C_{A_{N-1}}^{-1}}_{j,k}\varepsilon _{k}J_{j}\ ,  \label{hsol}
\end{equation}%
where $\varepsilon _{j}$ are signs, with $\varepsilon _{1}=1$ and $%
C_{A_{N-1}}$ is the Cartan matrix for $SU(N)$ (see \cite{gaugsksu(n)},
Proposition 2). Here, $J_{j}$ form a basis of the Cartan subalgebra of $SU(N)
$ defined as 
\begin{equation*}
J_{j}=i\left( E_{j,j}-E_{j+1,j+1}\right) \ .
\end{equation*}%
It follows that the energy density for the nuclear lasagna phase is given by 
\begin{align}
T_{00}=&-\frac{K}{2} L_\phi^2\text{Tr}\biggl[R_t^2+ \frac{1}{2}\biggl(\frac{%
R_r^2}{L_r^2}+\frac{R_\theta^2}{L_\theta^2}\biggl) +\frac{\lambda}{4}\biggl(%
\frac{F_{tr}^2}{L_r^2}+\frac{F_{t\theta}^2}{L_\theta^2}+\frac{1}{2}\frac{%
F_{r\theta}^2}{L_r^2\ L_\theta^2}\biggl)\biggl] \cr = & \frac K2 \|c\|^2
L_\phi^2\left( \frac {m^2}{L_\theta^2} +\frac {1}{8L_r^2} \frac {\|
v_\varepsilon\|^2}{\| c\|^2}+2 \frac {\sigma^2}{L_\phi^2} +\frac {\lambda m^2%
}{16L_r^2 L_\theta^2} + \frac {\lambda \sigma^2}{8L_\phi^2 L_r^2} \right) %
\cr &+\frac {K\lambda m^2\sigma^2}{ L_\theta^2} \sin^2{\left(\frac{r}{2}%
\right)}\left(\sum_{j=1}^{N-1} |c_j|^4+\sum_{j=1}^{N-2} |c_j|^2 |c_{j+1}|^2
\left(\frac 12-\frac 32\varepsilon_j \varepsilon_{j+1}\right) \right) \ ,
\label{4.4}
\end{align}
where we have denoted 
\begin{equation*}
\norm c^{2}=\sum_{j=1}^{N-1}|c_{j}|^{2}\ ,\qquad \norm {v_\varepsilon}^{2}=-%
\mathrm{Tr}\ v_{\varepsilon }^{2}\ .
\end{equation*}%
On the other hand the topological charge is 
\begin{equation}
B_{L}=2m\sigma \norm{c}^{2}\ .  \label{Blas}
\end{equation}%
In particular, we are interested in the case when $\varepsilon _{j}=1$ for
all $j$, so that 
\begin{gather*}
\norm{c}^{2}=\Lambda \norm {v_\varepsilon}^{2}=\frac{\Lambda }{12}%
N(N^{2}-1)\ , \\
\sum_{j=1}^{N-1}|c_{j}|^{4}-\sum_{j=1}^{N-2}|c_{j}|^{2}|c_{j+1}|^{2}=\frac{%
\Lambda }{2}\norm c^{2}\ ,
\end{gather*}%
where $\Lambda =\frac{1}{2}$ for odd $N$ and $\Lambda =2$ for even $N$ (see
Appendix B for the allowed values of the Baryonic charge as function of $N$).

\begin{figure}[h!]
\centering
\includegraphics[scale=.58]{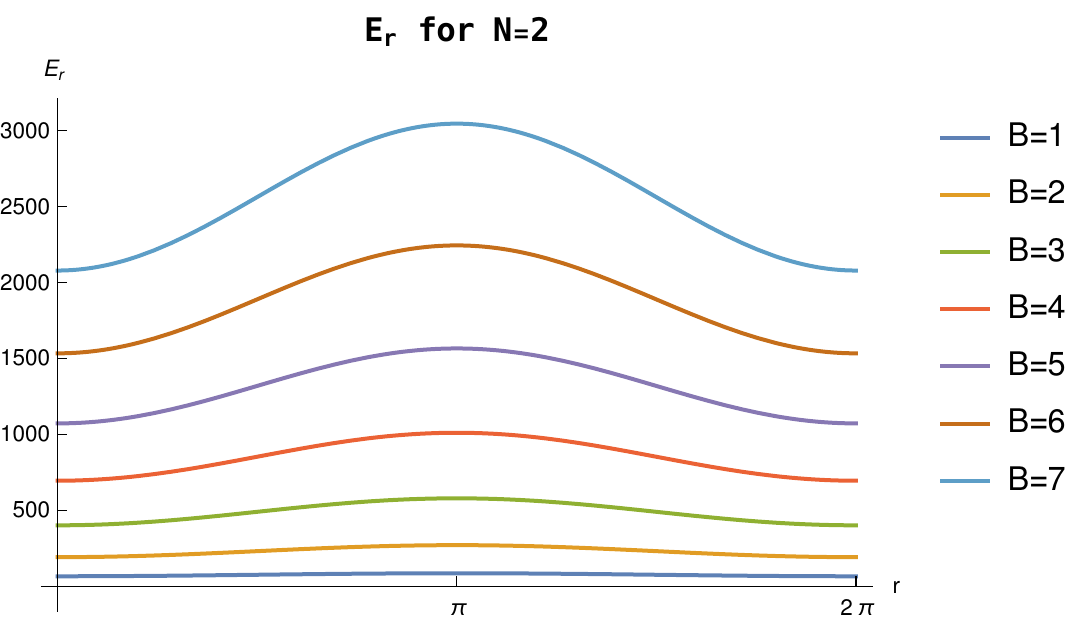} \qquad %
\includegraphics[scale=.58]{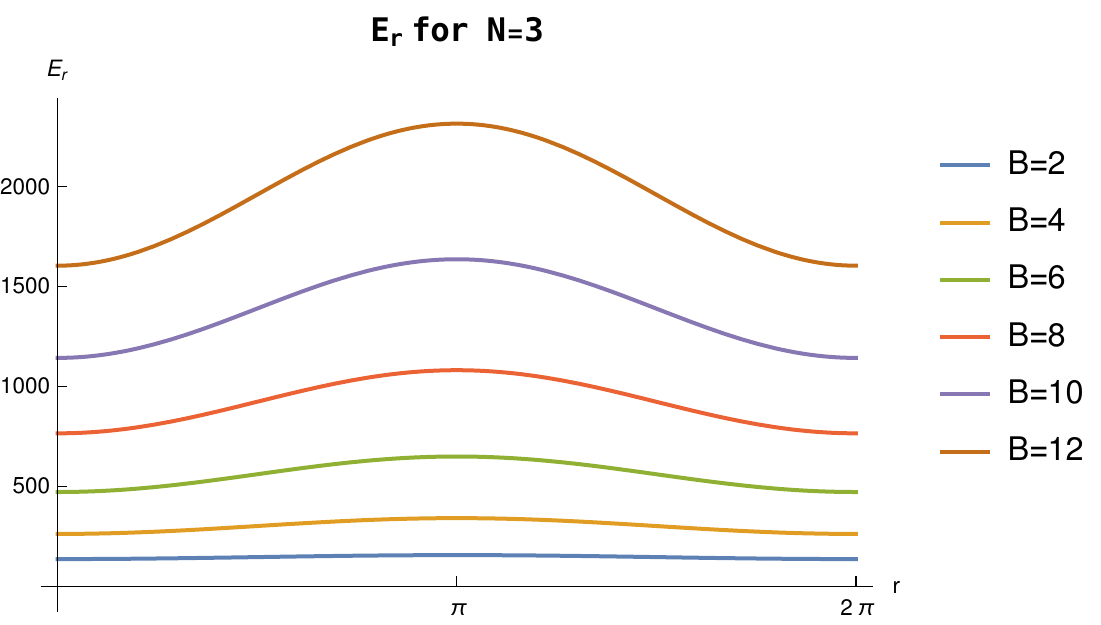}
\caption{Energy density in $r$ for the lasagna phase for $N=2,3$ and
different values of $B$. Here $L=1$ and $q=\frac{1}{2}$.}
\label{Fig5}
\end{figure}

Fig. \ref{Fig5} shows how the energy density changes for different values of
the topological charge and for fixed values of $N$. Although the behavior of
the energy density per Baryon is similar, the quotient $E/B$ does not depend
on $N$. 

\begin{figure}[h!]
\centering
\includegraphics[scale=.58]{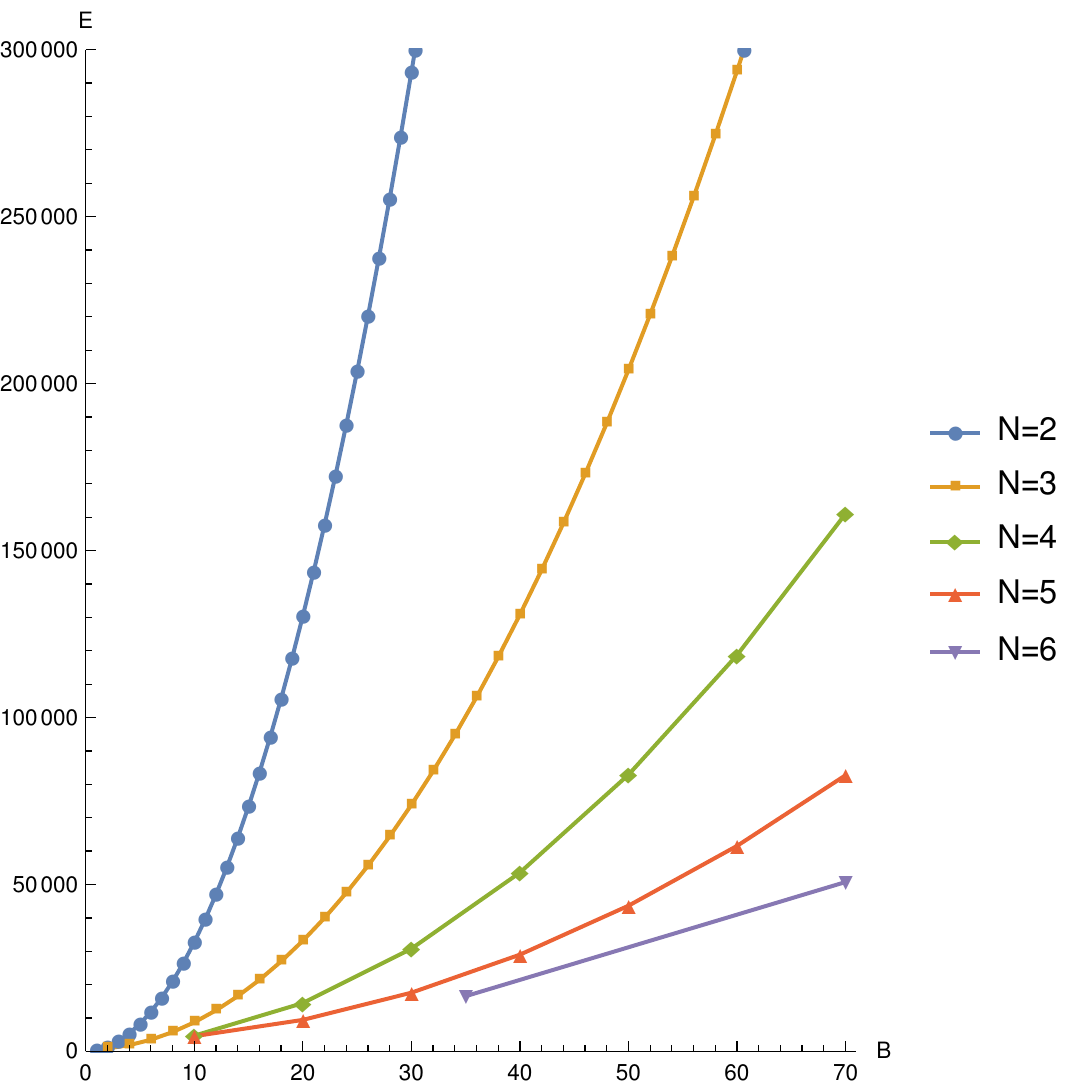} \qquad %
\includegraphics[scale=.58]{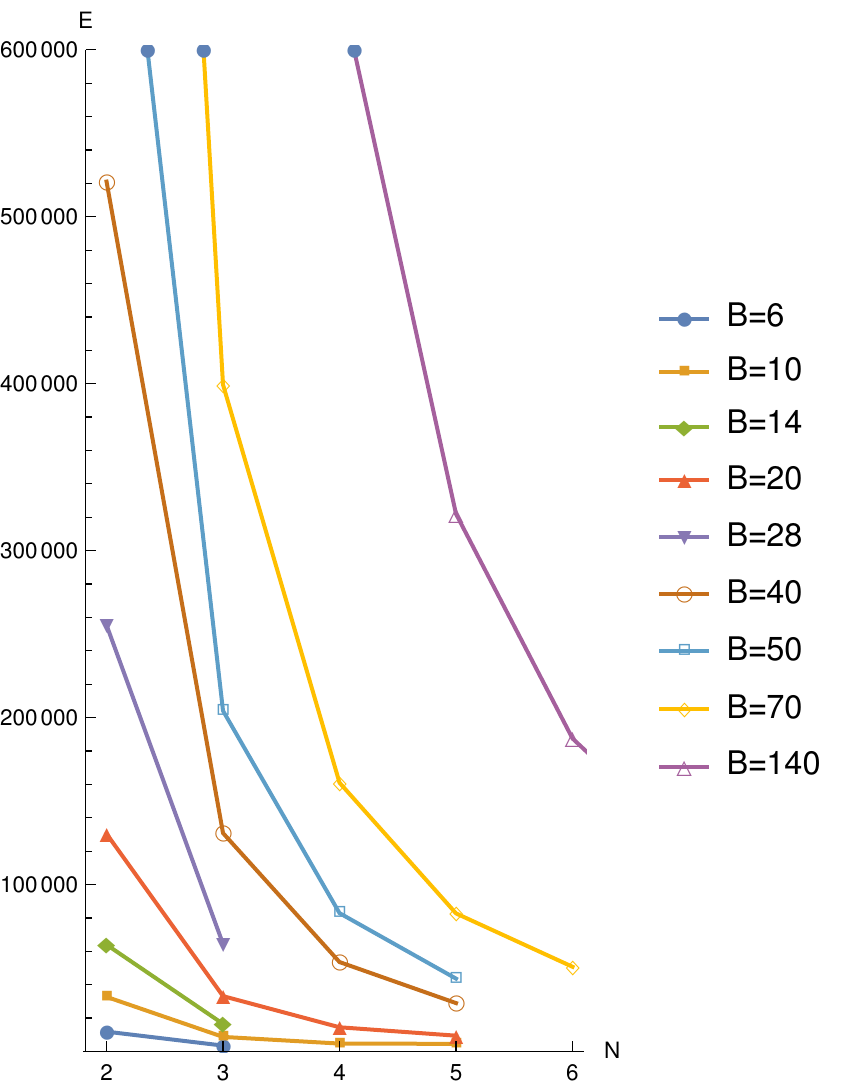}
\caption{Energy per Baryonic charge versus $B$ and energy versus $N$ for the
lasagna phase and for different values of $N$ and $B$, respectively. Here $%
L=1$ and $q=\frac{1}{2}$.}
\label{Fig6}
\end{figure}

In Fig. \ref{Fig6} one can see that the lasagna phase has the same behavior
as the spaghetti phase; its energy is an increasing quantity in $B$, and
that lasagna configurations belonging to theories with lower $N$ are more
energetic than those with larger $N$ values. 

\subsection{Comparing nuclear spaghetti and lasagna at finite Baryon density}

In this subsection, we will compare the energy densities of the lasagna and
spaghetti configurations.

First of all, it is important to note a key difference between the
expressions for the topological charge of the lasagna and the spaghetti
phases. Even though both configurations depends on the group dimension $N$
(see Appendix B), the spaghetti phase also depends on two more integers,
namely $n$ and $p$ (see Eq. \eqref{Beq}), while the lasagna depends only on
the integer $m$ (see Eq. \eqref{Blas}). Since we must compare configurations
with the same Baryonic charge, for a given value of $N$, in the case of the
spaghetti there will be many configurations that satisfy this requirement,
while for the lasagna phase there will be only one.

\begin{figure}[h]
\centering
\includegraphics[scale=.8]{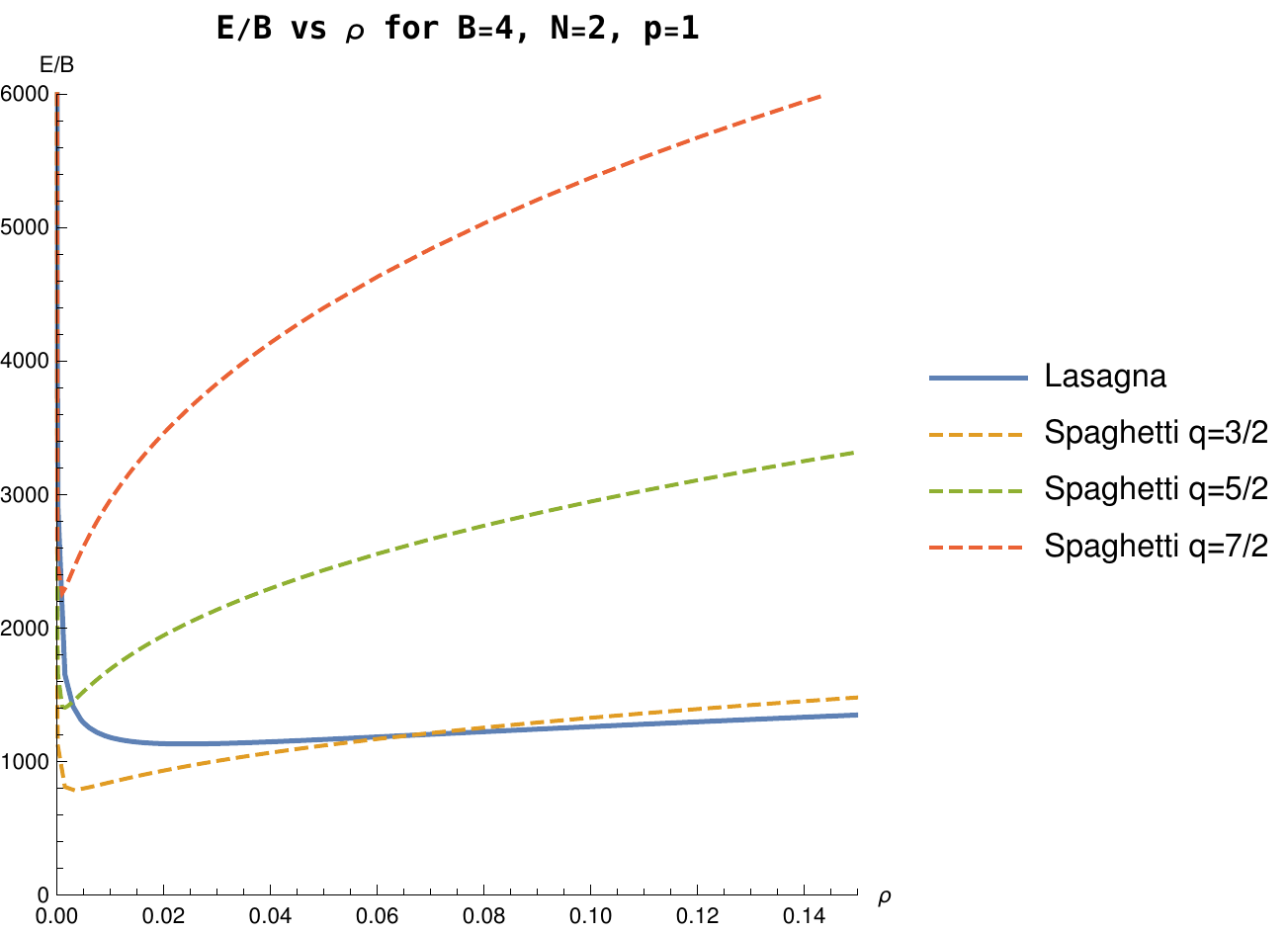}\newline
\vspace{.5cm} \includegraphics[scale=.8]{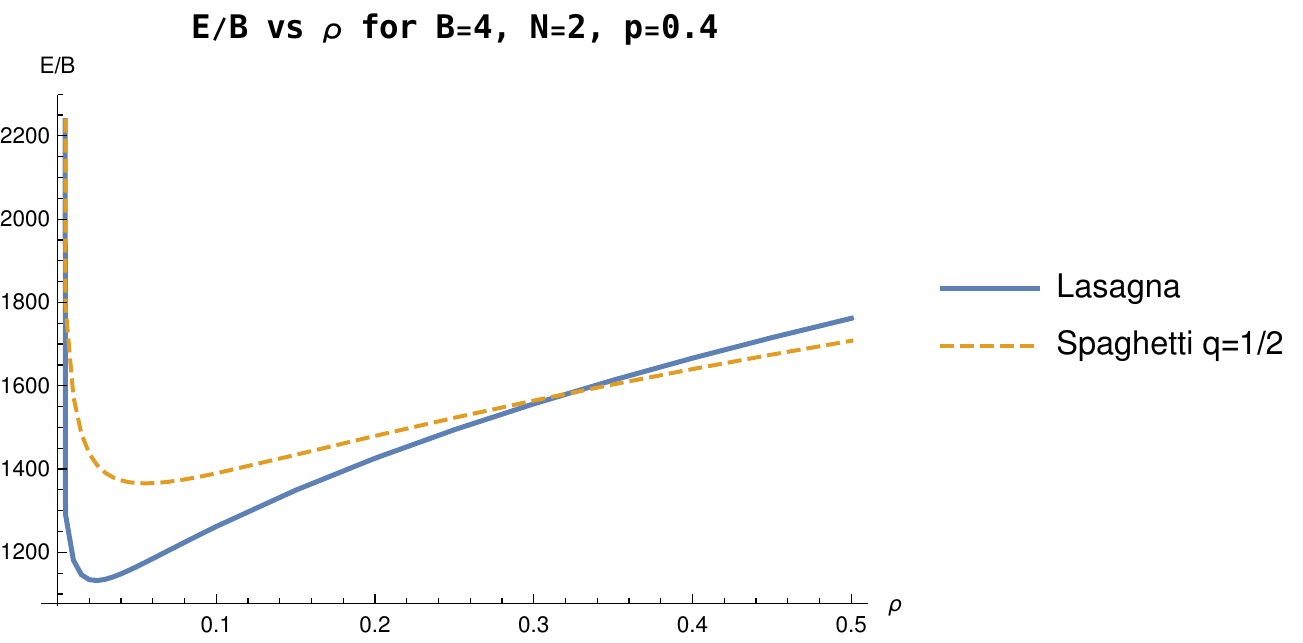}
\caption{Energy per Baryonic charge of the nuclear pasta phases as a
function of the density for different values of $q$. Above: For $p=1$, we
can see that below a certain value of $\rho$ the spaghetti phase has a lower energy
cost, while above this value the lasagna phase is the energetically
desirable. Note that the lasagna phase also exhibits the characteristic
\textquotedblleft u-shape\textquotedblright\ behavior shown in \protect\cite%
{pasta10}. Below: For $p=0.4$ and $q=\frac{1}{2}$, we see that for low densities the lasagna
phase has a lower energy cost, while for high density the spaghetti phase is
the energetically favored.}
\label{Fig8}
\end{figure}

According to the expressions for the energy of the spaghetti and lasagna
phases in Eqs. \eqref{3.14} and \eqref{4.4}, respectively, one can see that when the
total energy (for fixed values of the Baryonic charge and fixed flavor
number) is computed as a function of the density, at high
density (but still well within the range of validity of the Skyrme model)
the lasagna configurations are energetically favored while at low density
the spaghetti configurations are favored (see Fig. \ref{Fig8}).

It is also important to note that the $q$ parameter present in the
spaghetti-like configurations plays a very important role
as $2qn$ represents the number of spaghetti in the box. 

From Fig. \ref{Fig8} it can be
seen that the $q$ parameter plays a very important role, as can also be seen from the plots. Indeed, the plot in Fig. %
\ref{Fig8} confirms that at high density lasagna configurations are
energetically favoured while at low density spaghetti configurations are
energetically favoured (the density at which one type of
configuration overcomes the other depends on the parameter $q$).

\subsection{A comment on the large $N$ limit}

Although the large $N$ limit of the $SU(N)$-Skyrme model is not directly
relevant in the phenomenology of nuclear pasta, it has a great theoretical
interest since it is connected with the Veneziano limit of the large $N_{c}$
expansion of QCD, $N_{c}$ being the number of colors, in which also the
flavor number $N$ goes to infinity keeping constant $N/N_{c}$ (see \cite%
{largeN1}, \cite{largeN2} and \cite{largeN3}). Here below we present a
result that is of interest in this context (in which it is always an
important achievement to show that physically meaningful quantities do
possess a smooth large $N$ limit). The energy per Baryon, $g(B,N,L)=E/B$,
for the spaghetti and lasagna phases are given respectively by 
\begin{align}
g_{\text{S}}(B,N,L)=&\frac{L^{3}K(2\pi )^{2}}{8np}\left[
4\xi_{1}(q^{2}+p^{2})+n^{2}\xi_{4}+\frac{\lambda }{L^{2}}\left(
n^{2}\xi_{3}(q^{2}+p^{2})+4\xi_{2}q^{2}p^{2}\right) \right] \ ,  \label{gS} \\
g_{\text{L}}(B,N,L)=&\frac{L^{3}K (2\pi )^{3}}{2 \sigma m}\left[ m^{2}+\frac{%
1}{8\Lambda }+2\sigma ^{2}+\frac{\lambda }{L^{2}}\left( \frac{m^{2}}{16}+%
\frac{\sigma ^{2}}{8}+\frac{\Lambda m^{2}\sigma ^{2}}{2}\right) \right] \ ,
\label{gL}
\end{align}%
where we have defined the integrals 
\begin{align}
	\xi_{1}& =\int_{0}^{2\pi }\sin ^{2}\left( \frac{\chi (r)}{2}\right) dr\ , \\
	\xi_{2}& =\int_{0}^{2\pi }\sin ^{4}\left( \frac{\chi (r)}{2}\right) dr\ , \\
	\xi_{3}& =\frac{1}{n^{2}}\int_{0}^{2\pi }\sin ^{2}\left( \frac{\chi (r)}{2}%
	\right) \eta (r)^{2}dr\ , \\
	\xi_{4}& =\frac{1}{n^{2}}\int_{0}^{2\pi }\eta (r)^{2}dr\ ,
\end{align}%
and $\eta (r)$ is defined in Eq. (\ref{qua}) (here, $\xi_j$ do not scale with $N$). A fact that can be seen explicitly from our analytical construction is that
the above quantities do not depend on $N$ in a \textquotedblleft significant
way\textquotedblright . In particular, the energy per Baryon for the
spaghetti case in Eq. \eqref{gS} does not depends on $N$ at all, while for
the lasagna case in Eq. \eqref{gL} the dependence is stored in the
definition of $\Lambda $ and $\sigma $ and it takes different finite values
depending on whether $N$ is even or odd. Consequently, we have shown that
the energy per Baryon $g_{\text{S}}(B,N,L)$ in the spaghetti case has a
smooth well defined large $N$ limit. In the lasagna case, $g_{\text{L}%
}(B,N,L)$ also possesses a well defined large $N$ limit provided we treat
separately the case in which $N$ is large and even and the case in which $N$
is large and odd.

\subsection{Subleading corrections}

In previous sections we mentioned that the configurations shown in this work
are also solutions of the generalized Skyrme model including higher order
corrections in the 't Hooft limit. It is worth to emphasize that this claim
could appear to be quite unrealistic due to the extremely complex nature of
the subleading corrections to the Skyrme model arising in the 't Hooft
expansion \cite{subleading1}. Indeed, until very recently not only there was
no analytic and topologically non-trivial solution of the Skyrme field
equations modified by the corrections presented in \cite{subleading1}, but
such corrections were neglected also in the numerical analysis due to their
highly non-linear character.

In fact, in \cite{crystal3} it was shown explicitly how such corrections
modify the analytic spaghetti configurations for the $SU(2)$ group. In order
to expand these results, in this subsection we will explicitly show a
similar result for the lasagna phase in the cases of most physical interest:
namely, for the internal groups with $N=2$ and $N=3$.

In four dimensions, we have that the low energy limit of QCD at leading
order in the 't Hooft expansion can be described by the following action: 
\begin{gather}
\hat{I}=I+\int d^{4}x\sqrt{-g}\mathcal{L}_{\text{corr}} \ ,  \label{sky1}
\end{gather}%
where $I$ corresponds to the Skyrme action defined above in Eq. \eqref{I}
and the terms $\mathcal{L}_{\text{corr}}$ represents the possible subleading
corrections to the Skyrme model which can be computed, in principle, using
either Chiral Perturbation Theory (see \cite{Scherer:2002tk} and references
therein) or the large \textbf{N} expansion \cite{largeN1}, \cite{largeN3}.
The expected corrections have the following generic form 
\begin{align}
\mathcal{L}_{6}=& \frac{c_{6}}{96}\text{Tr}\left[ F_{\mu }{}^{\nu }F_{\nu
}{}^{\rho }F_{\rho }{}^{\mu }\right] \ ,  \notag \\
\mathcal{L}_{8}=& -\frac{c_{8}}{256}\biggl(\text{Tr}\left[ F_{\mu }{}^{\nu
}F_{\nu }{}^{\rho }F_{\rho }{}^{\sigma }F_{\sigma }{}^{\mu }\right] -\text{Tr%
}\left[ \{F_{\mu }{}^{\nu },F_{\rho }{}^{\sigma }\}F_{\nu }{}^{\rho
}F_{\sigma }{}^{\mu }\right] \biggl)\ ,  \label{Lcorr}
\end{align}%
and so on \cite{subleading1}, where the $c_{p}$ ($p\geq 6$) are subleading
with respect to $K$ and $\lambda $.

The field equations of the model are obtained varying the action in Eq. %
\eqref{sky1} w.r.t. the $U$ field. To perform this derivation it is useful
to keep in mind the following relation 
\begin{equation*}
\delta _{U}R_{\mu }=[R_{\mu },U^{-1}\delta U]+\partial _{\mu }(U^{-1}\delta
U)\ ,
\end{equation*}%
where $\delta _{U}$ denotes variation w.r.t the $U$ field. From the above,
the field equations of the generalized Skyrme model turns out to be 
\begin{align}
\frac{K}{2}\biggl(\partial ^{\mu }R_{\mu }+\frac{\lambda }{4}\partial ^{\mu
}[R^{\nu },F_{\mu \nu }]\biggl)+3c_{6}[R_{\mu },\partial _{\nu }[F^{\rho \nu
},F_{\rho }{}^{\mu }]]&   \notag \\
+4c_{8}\biggl[R_{\mu },\partial _{\nu }\biggl(F^{\nu \rho }F_{\rho \sigma
}F^{\sigma \mu }+F^{\mu \rho }F_{\rho \sigma }F^{\nu \sigma }+\{F_{\rho
\sigma },\{F^{\mu \rho },F^{\nu \sigma }\}\}\biggl)\biggl]& \ =\ 0\ .
\label{sessea0}
\end{align}%
On the other hand the energy-momentum tensor of the theory including the
subleading corrections is 
\begin{equation}
T_{\mu \nu }\ =\ T_{\mu \nu }^{\text{Sk}}+T_{\mu \nu }^{(6)}+T_{\mu \nu
}^{(8)}\ ,  \label{Tmunu}
\end{equation}%
where a direct computation reveals that 
\begin{align*}
T_{\mu \nu }^{(6)}=& -\frac{c_{6}}{16}\text{Tr}\biggl(g^{\alpha \gamma
}g^{\beta \rho }F_{\mu \alpha }F_{\nu \beta }F_{\gamma \rho }-\frac{1}{6}%
g_{\mu \nu }F_{\alpha }{}^{\beta }F_{\beta }{}^{\rho }F_{\rho }{}^{\alpha }%
\biggl)\ , \\
T_{\mu \nu }^{(8)}=& \ \frac{c_{8}}{32}\text{Tr}\biggl(g^{\alpha \rho
}g^{\beta \gamma }g^{\delta \lambda }F_{\alpha \mu }F_{\nu \beta }F_{\gamma
\delta }F_{\lambda \rho }+\frac{1}{2}\{F_{\mu \alpha },F_{\lambda \rho
}\}\{F_{\beta \nu },F_{\gamma \delta }\}g^{\alpha \gamma }g^{\beta \rho
}g^{\delta \lambda } \\
& -\frac{1}{8}g_{\mu \nu }(F_{\alpha }{}^{\beta }F_{\beta }{}^{\rho }F_{\rho
}{}^{\sigma }F_{\sigma }{}^{\alpha }-\{F_{\alpha }{}^{\beta },F_{\rho
}{}^{\sigma }\}F_{\beta }{}^{\rho }F_{\sigma }{}^{\alpha })\biggl)\ .
\end{align*}%
Now, considering the very same ansatz for the lasagna phase given in Eqs. %
\eqref{U2} and \eqref{U2b} we obtain that the complete set of coupled Skyrme
field equations reduce the single integrable ODE for the $SU(2)$ and $SU(3)$
cases, respectively 
\begin{gather}
\biggl(16KL_{r}^{2}L_{\theta }^{2}(4L_{\theta }^{2}+m^{2}\lambda
)-3c_{8}m^{4}h^{\prime 2}\biggl)h^{\prime \prime }=0\ , \\
\biggl(4KL_{r}^{2}L_{\theta }^{2}(4L_{\theta }^{2}+m^{2}\lambda
)-3c_{8}m^{4}h^{\prime 2}\biggl)h^{\prime \prime }=0\ .
\end{gather}%
We write the field equations in this way in order to make clear that the
inclusion of these corrections keep intact the nice structure of the field
equations. It is also quite manifest that the lasagna solutions presented in
this work, when the profile $h(r)$ is a linear function as in Eq. %
\eqref{hsol}, also satisfy the field equations of this generalized model.
This is a quite remarkable result in itself since, until very recently, such
corrections were not even included in numerical analysis while with the
present approach can be studied analytically.

Now, despite the fact that the high order corrections do not affect the
structure of the field equations, these terms do play a rol in the physical
properties of these configurations. In particular, the energy density of
this phase will be modified according to 
\begin{equation*}
T_{00}^{(6)}=\frac{1}{128}\frac{c_{6}m^{2}}{L_{r}^{2}L_{\theta }^{2}L_{\phi
}^{2}}\sin ^{2}(\frac{r}{2})\ ,\ \ T_{00}^{(8)}=\frac{1}{2^{14}}\frac{%
c_{8}m^{2}}{L_{r}^{4}L_{\theta }^{4}L_{\phi }^{2}}\biggl(8L_{\theta
}^{2}+(8L_{r}^{2}-L_{\phi }^{2})m^{2}+4(L_{\theta }^{2}-2L_{r}^{2}m^{2})\cos 
(r)\biggl)\ ,
\end{equation*}%
for $SU(2)$ and 
\begin{equation*}
T_{00}^{(6)}=\frac{1}{32}\frac{c_{6}m^{2}}{L_{r}^{2}L_{\theta }^{2}L_{\phi
}^{2}}\sin ^{2}(\frac{r}{2})\ ,\ \ T_{00}^{(8)}=\frac{1}{2^{10}}\frac{%
c_{8}m^{2}}{L_{r}^{4}L_{\theta }^{4}L_{\phi }^{2}}\biggl(8L_{\theta
}^{2}+(8L_{r}^{2}-L_{\phi }^{2})m^{2}+4(L_{\theta }^{2}-2L_{r}^{2}m^{2})\cos 
(r)\biggl)\ ,
\end{equation*}%
for the $SU(3)$ case. 

It is worth to note that the fact that the $N=3$ solution has lower energy
in the ground state than the $N=2$ solution is an artifact of the model
which can be taken into account including subleading corrections in the 't
Hooft large-$N_{c}$ expansion. However, although in the present approach
these subleading corrections do not spoil the integrability of the field
equations, such corrections certainly make the analytic formulas for the
energy density and other relevant physical quantities considerably more
cumbersome (see also \cite{crystal3}). Therefore, we decided to consider explicitly
only the Skyrme model since it already captures a great deal of novel
physical information on the nuclear pasta phase keeping, at the same time,
the analytic formulas readable. We plan to work on the effects of these
subleading corrections in a future publication.

\subsection{Isospin chemical potential} 

We have shown in previous sections that the inclusion of a suitable
time-dependence in the ans\"{a}tze, both for lasagna and spaghetti phases
(see Eqs. \eqref{ansatz} and \eqref{U2b}), is one of the key ingredients
that allows the field equations to be considerably reduced, leading to a
single integrable ODE equation for the profiles. This time-dependence offers
a nice short-cut to estimate the ``classical Isospin" of the configurations
analyzed in the present paper (a relevant question is whether or not the
classical Isospin is large when the Baryonic charge is large). In
particular, one may evaluate the ``cost" of removing such time-dependence.
Such a cost is related to the internal Isospin symmetry of the theory. This
is like trying to estimate the angular momentum of a spinning top by
evaluating the cost to make the spinning top to stop spinning. In the
present case, the time-dependence of the configurations can be removed from
the ans\"{a}tze by introducing a  Isospin chemical potential; then the
isospin chemical potential needed to remove such time-dependence is a
measure of the classical Isospin of the present configurations. We will
see how this works for the simplest $SU(2)$ case, where the generators are $%
t_{j}=i\sigma _{j}$, being $\sigma _{j}$ the Pauli matrices (higher $SU(N)$
behave in a similar way).

As it is well known, the effects of the Isospin chemical potential can be
taken into account by using the following covariant derivative 
\begin{equation}
\nabla _{\mu }\rightarrow D_{\mu }=\nabla _{\mu }+\bar{\mu}[t_{3},\cdot
]\delta _{\mu 0}\ .  \label{Dmu}
\end{equation}%
Now, we will use exactly the same ansatz as before in the spaghetti $SU(2)$
case, \textit{but this time without the time dependence}: 
\begin{gather*}
U=e^{\chi (x)\,(\vec{n}\cdot \vec{t})}\ , \\
\vec{n}=(\sin {\Theta }\sin \Phi ,\sin \Theta \cos \Phi ,\cos \Theta )\ ,
\end{gather*}%
where 
\begin{gather*}
\chi =\chi \left( r\right) \ ,\quad \Theta =q\theta \ ,\quad \Phi =p\phi \ ,
\\
q=\frac{1}{2}(2v+1)\ ,\quad p,v\in \mathbb{N}\ ,\quad p\neq 0\ ,
\end{gather*}%
together with the introduction of the Isospin chemical potential in Eq. %
\eqref{Dmu} in the theory. One can check directly that the complete set of
Skyrme equations can still be reduced to the same ODE for the profile $\chi
\left( r\right) $ in the case of the spaghetti phase in Eq. \eqref{Eqchi} 
\textit{only provided the Isospin chemical potential for the spaghetti phase
is given by} 
\begin{equation}
\bar{\mu}_{\text{S}}=\frac{p}{L_{\phi }}\ .
\end{equation}%
In other word, the cost to eliminate the time-dependence is to introduce an
Isospin chemical potential which is large when the Baryonic charge of the
spaghetti is large. Something similar happens in the case of the lasagna
phase. Let us consider the ansatz in terms of the Euler angles \textit{but
without the time-dependence} for the $SU(2)$ case: 
\begin{equation*}
U_{L}=e^{\Phi t_{3}}e^{Ht_{2}}e^{\Theta t_{3}}\ ,
\end{equation*}%
where 
\begin{equation*}
\Phi =p\phi \ ,\quad H=h(r)\ ,\quad \Theta =m\theta \ ,\qquad p,m\in \mathbb{%
N}\ .
\end{equation*}%
Let us introduce the Isospin chemical potential, demanding that the profile $%
h(r)$ should be the same as before. Then, as in the spaghetti case, the
Skyrme field equations with chemical potential can still be satisfied by the
very same profile $h(r)$ \textit{provided we fix the Isospin chemical
potential as} 
\begin{equation}
\bar{\mu}_{\text{L}}=\frac{pm}{(p^{2}L_{\phi }^{2}+m^{2}L_{\theta }^{2})^{%
\frac{1}{2}}}\ .
\end{equation}%
At this point it is important to remember that in the $SU(2)$ case the
lasagna and spaghetti type solutions have the following values for the
topological charges 
\begin{equation*}
B_{\text{S}}=np\ ,\qquad B_{\text{L}}=mp\ ,
\end{equation*}%
see \cite{56c} and \cite{crystal1} for more details. Therefore, from the
above computations it can be seen that the price to pay to eliminate the
time-dependence from the configurations discussed in the previous sections
is to introduce an Isospin chemical potential which grows as the Baryonic
charge of these configurations grows. These arguments show that the
``classical Isospin" of configurations with high Baryonic charge is large.
Finally, it is important to point out that the large Isospin case corresponds to either neutron rich or proton rich
matter and due to Coulomb effects (not taken into account in this model), the
neutron rich solution is preferred. This is very nice for the physics relevant
for neutron stars\footnote{We thanks the Referee for this valuable comment.}.

%%%%%%%%%%%%%%%%%%%%%%%%%%%%%%%%%%%%%%%%%%%%%%%%%%%%%%%%%%%%%%%%%%%%%

\section{Conclusions and perspectives}

%%%%%%%%%%%%%%%%%%%%%%%%%%%%%%%%%%%%%%%%%%%%%%%%%%%%%%%%%%%%%%%%%%%%%

In this manuscript, by combining the strategy of \cite{crystal1}, \cite%
{crystal2}, \cite{crystal3}, \cite{crystal4}, \cite{firstube} and \cite%
{gaugsksu(n)} with the generalization of the Euler angles to $SU(N)$ of \cite%
{euler1}, \cite{euler2} and \cite{euler3}, we have constructed
multi-Baryonic solutions living at finite Baryon density in the $SU(N)$%
-Skyrme model for generic values of the number of flavors and with arbitrary
values of the topological charge. The energy density of these configurations
is concentrated either in tube-shaped regions (suitable to describe the
nuclear spaghetti phase) or within Baryonic layers of finite width (suitable
to describe the nuclear lasagna phase). To the best of author's knowledge,
this is the first systematic tool to construct analytic topologically
non-trivial solutions of such complexity in the $SU(N)$-Skyrme model for
generic $N$. The physical interest of the present configurations is
confirmed, for instance, by the fact that our construction shows explicitly
that, at lower densities, configurations with $N=2$ light flavors are
favoured while, at higher densities, configurations with $N=3$ are favoured.
This is a quite non-trivial test of the present approach as a serious
analytic candidate to describe the nuclear pasta phase.

The importance of these analytic results arises from the fact that they allow to compare explicitly a relevant physical properties of nuclear lasagna and nuclear spaghetti.
For instance, one can see that for high density (but still well within the range of
validity of the Skyrme model) the lasagna configurations are favored while
at low density the spaghetti configurations are favored. Since the
physical properties (such as thermal and electric conductivities) of lasagna
and spaghetti phases are very different (see \cite{pasta10} and references
therein), our results can have interesting
phenomenological implications in high density particles physics, neutron
stars and so on. We have also discussed relevant physical quantities (such
as the energy per Baryon) which have a smooth large $N$ limit. We hope to
come back on the intriguing observable effects related to the present
results in a future publication.

\subsection*{Acknowledgments}

We are grateful to an anonymous referee, whose comments helped us in severely  improve our manuscript.\\
F. C. has been funded by Fondecyt Grant 1200022. M. L. is funded by FONDECYT
post-doctoral Grant 3190873. A. V. is funded by FONDECYT post-doctoral Grant
3200884. The Centro de Estudios Cient\'{\i}ficos (CECs) is funded by the
Chilean Government through the Centers of Excellence Base Financing Program
of ANID.

\section*{Appendix A: Mathematical tools}

In this Appendix we want to recall some main facts about the groups $SU(N)$.
Recall that $SU(N)$ is the compact simply connected real group consisting on
the set of $N\times N$ complex matrices $U$ satisfying the relations 
\begin{align}
U^\dagger U \ =& \ \mathbb{I} \ ,  \label{unitarity} \\
\det U \ = & \ 1 \ ,  \label{speciality}
\end{align}
where $\mathbb{I}$ is the $N\times N$ identity matrix. Its corresponding Lie
algebra is the real vector space of traceless anti-Hermitian matrices 
\begin{equation}
A=-A^\dagger \ .
\end{equation}
To see this, consider a curve $U(t)\in SU(N)$, such that $U(0)=\mathbb{I}$.
By definition, the element of the Lie algebra are the matrices of the form 
\begin{equation*}
A=\frac {dU}{dt}(0) \ .
\end{equation*}
Deriving the relation 
\begin{equation*}
U(t)^\dagger U(t)=\mathbb{I} \ ,
\end{equation*}
we get to 
\begin{equation*}
\frac {dU^\dagger}{dt}(t) U(t)+U(t)^\dagger \frac {dU}{dt}(t)=\mathbb{O} \ ,
\end{equation*}
where $\mathbb{O}$ is the zero $N\times N$ matrix. Setting $t=0$ gives $%
A^\dagger+A=\mathbb{O}$, which show anti-Hermitian-ness. Reality follows
from the fact that linear combinations of anti-Hermitian matrices is
anti-Hermitian if and only if the coefficients are real. The traceless
condition follows from the speciality condition in Eq. \eqref{speciality}.
Indeed, recall that developing the determinant of a matrix $X$ with respect
to the $j $-th row one has 
\begin{equation*}
\det X=\sum_k x_{jk} \mathrm{cof}_{jk}(X) \ ,
\end{equation*}
where $x_{ij}$ are the matrix elements and $\mathrm{cof}_{jk}(X)$ is $%
(-1)^{j+k}$ times the determinant of the reduced matrix after eliminating
the $j$-th row and the $k$-th column. From this we get to 
\begin{equation*}
\frac {d}{dx_{jk}} \det X=\mathrm{cof}_{jk}(X) \ .
\end{equation*}
Therefore, deriving Eq. \eqref{speciality} w.r.t. $t$ in $t=0$ gives (for $%
u_{jk} $ the matrix elements of $U$) 
\begin{align}
0=\left. \frac d{dt} \det U\right|_{t=0}=\left. \sum_{j,k} \frac {d (\det
U(t))}{du_{jk}(t)} \frac {du_{jk}}{dt}\right|_{t=0}=\sum_{j,k} \mathrm{cof}%
_{jk}(U(0)) A_{jk} \ .
\end{align}
On the other hand, using that $U(0)=\mathbb{I}$ and $\mathrm{cof}%
_{jk}(I)=\delta_{jk}$, we lead to 
\begin{align}
0=\sum_{j,k} \delta_{jk}A_{jk}=\sum_j A_{jj}= \mathrm{Trace}(A) \ .
\end{align}
In particular, $\text{Lie}(SU(N))\equiv su(N)$ has dimension $(N^2-1)$.

Recall that one can back from the algebra to the group by mean of the
exponential map 
\begin{equation}
\exp: Lie(G) \longrightarrow G \ , \qquad A\longmapsto \exp A \ ,
\end{equation}
which for a matrix group is 
\begin{equation*}
\exp A= \mathbb{I}+\sum_{n=1}^\infty \frac 1{n!} A^n \ .
\end{equation*}
In particular, for a compact Lie group $\exp$ is surjective, so any group
element can be written as an exponential. Therefore, fixing a basis $t_j$
for the algebra, any element of the group can be written as 
\begin{equation*}
g=\exp (\sum_j x^jt_j) \ , \quad\ x^j\in \mathbb{R} \ .
\end{equation*}
For $SU(N)$ a convenient basis is given by the anti-Hermitian Gell-Mann
matrices constructed as follows (see \cite{euler1}): 
\begin{align}
t_{a^2-1}&= \sqrt {\frac 2{a(a-1)}} \left(\sum_{j=1}^{a-1} E_{j,j}-(a-1)
E_{a,a}\right) \ , \qquad a=2,\ldots, N, \\
t_1&=-i (E_{1,2}+E_{2,1}) \ , \qquad t_2= E_{1,2}-E_{2,1}, \\
t_{a^2-1+2j}&=E_{j,a+1}-E_{a+1,j} \ , \qquad j=1,\ldots, a, \quad
a=2,\ldots, N-1, \\
t_{a^2-2+2j}&=-i(E_{j,a+1}+E_{a+1,j}) \ , \qquad j=1,\ldots, a, \quad
a=2,\ldots, N-1.
\end{align}
Here $E_{j,k}$ is the $N\times N$ matrix whose only non zero element is $%
(E_{j,k})_{jk}=1$. The Gell-Mann matrices $t_{A}$, $A=1,\ldots,(N^2-1)$, are
normalized, so that 
\begin{equation*}
-\frac 12 \mathrm{Tr} (t_A t_B)=\delta_{AB} \ .
\end{equation*}
There are $(N-1)$ diagonal matrices according to the fact that $SU(N)$ has
rank $(N-1)$. We can also determine the maximal embedding of $SU(2)$ inside $%
SU(N)$. This is obtained by identifying $N$ as the dimension of an
irreducible representation of $SU(2)$. This means that we have to consider
the representation of spin $j=(N-1)/2$. This is standard and given by the
matrices in Eqs. (\ref{T1}), (\ref{T2}) and (\ref{T3}). In particular, we
see that we get a Bosonic representation for $N$ odd and a Fermionic
representation for $N$ even.

\section*{Appendix B: Topological charge for the nuclear phases}

The allowed values of the topological charge for the spaghetti and lasagna
phases depends on the flavor number $N$, that are explicitly shown in Eqs. %
\eqref{Beq} and \eqref{Blas}. This determines that for a given value of $N$
the topological charge cannot take arbitrary values. In Table \ref{TableB},
we show the allowed values of the topological charge from $N=2$ to $N=10$. 
\begin{table}[h!]
\begin{center}
\begin{tabular}{|c|l|l|}
\hline
\multicolumn{3}{|c|}{\textbf{Baryonic charge as $B=B(N)$}} \\ \hline
$N$ & $\quad Spaghetti$ & $\quad Lasagna$ \\ \hline
$2$ & $n p=\{1,2,3,4,...\}$ & $m=\{1,2,3,4,...\}$ \\ 
$3$ & $4 n p=\{4,8,12,16,...\}$ & $2m=\{2,4,6,8,...\}$ \\ 
$4$ & $10n p=\{10,20,30,40,...\}$ & $10m=\{10,20,30,40,...\}$ \\ 
$5$ & $20 np=\{20,40,60,80,...\}$ & $10m=\{10,20,30,40,...\}$ \\ 
$6$ & $35n p=\{35,70,105,140, ..\}$ & $35m=\{35,70,105,140, ..\}$ \\ 
$7$ & $56 n p=\{56,112,168,280,...\}$ & $28m=\{28,56,84,112,...\}$ \\ 
$8$ & $84 np=\{84,168,252,336,... \}$ & $84m=\{84,168,252,336,... \}$ \\ 
$9$ & $120 np=\{120,240,360,480,... \}$ & $60m=\{60,120,180,240,... \}$ \\ 
$10$ & $165 np=\{165,330,495,660,... \}$ & $165m=\{165,330,495,660,... \}$
\\ \hline
\end{tabular}%
\end{center}
\caption{Baryonic charge of the nuclear spaghetti and lasagna phases for
different values of $N$.}
\label{TableB}
\end{table}

\end{document}